%
%

\input harvmac

\input epsf
\newcount\figno
\figno=0
\def\fig#1#2#3{
\par\begingroup\parindent=0pt\leftskip=1cm\rightskip=1cm\parindent=0pt
\baselineskip=11pt
\global\advance\figno by 1
\midinsert
\epsfxsize=#3
\centerline{\epsfbox{#2}}
\vskip 12pt
{\bf Figure \the\figno:} #1\par
\endinsert\endgroup\par
}
\def\figlabel#1{\xdef#1{\the\figno}}
\def\encadremath#1{\vbox{\hrule\hbox{\vrule\kern8pt\vbox{\kern8pt
\hbox{$\displaystyle #1$}\kern8pt}
\kern8pt\vrule}\hrule}}

\batchmode
  \font\bbbfont=msbm10
\errorstopmode
\newif\ifamsf\amsftrue
\ifx\bbbfont\nullfont
  \amsffalse
\fi
\ifamsf
\def\IR{\hbox{\bbbfont R}}
\def\IZ{\hbox{\bbbfont Z}}
\def\IF{\hbox{\bbbfont F}}
\def\IP{\hbox{\bbbfont P}}
\else
\def\IR{\relax{\rm I\kern-.18em R}}
\def\IZ{\relax\ifmmode\hbox{Z\kern-.4em Z}\else{Z\kern-.4em Z}\fi}
\def\IF{\relax{\rm I\kern-.18em F}}
\def\IP{\relax{\rm I\kern-.18em P}}
\fi


\def\FI{Fayet-Iliopoulos}


\def\np#1#2#3{Nucl. Phys. {\bf B#1} (#2) #3}
\def\pl#1#2#3{Phys. Lett. {\bf #1B} (#2) #3}

\def\cmp#1#2#3{Comm. Math. Phys. {\bf #1} (#2) #3}

\def\cqg#1#2#3{Class. Quant. Grav. {\bf #1} (#2) #3}

\lref\hw{A. Hanany and E. Witten, ``Type IIB Superstrings, BPS
Monopoles, and Three Dimensional Gauge Dynamics,'' hep-th/9611230.}%

\lref\ahw{I. Affleck, J. Harvey and E. Witten, ``Instantons and
(Super-)Symmetry Breaking in (2+1) Dimensions,'' \np{206}{1982}{413}.}%

\lref\swthree{N. Seiberg and E. Witten, ``Gauge Dynamics and
Compactification to Three Dimensions,'' hep-th/9607163.}%

\lref\katzvafa{S. Katz and C. Vafa, ``Geometric Engineering of $N=1$
Quantum Field Theories,'' hep-th/9611090.}%

\lref\hk{A. Hanany and I. R. Klebanov, ``On Tensionless Strings in
(3+1) Dimensions,'' hep-th/9606136.}%

\lref\weinberg{E. Weinberg, ``Fundamental Monopoles and Multimonopole
Solutions for Arbitrary Simple Gauge Groups,'' \np{167}{1980}{500}.}%

\lref\callias{C. Callias, ``Index Theorems on Open Spaces,''
\cmp{62}{1978}{213}.}%

\lref\seibergfive{N. Seiberg, ``Five Dimensional SUSY Field Theories,
Non-Trivial Fixed Points and String Dynamics,'' hep-th/9608111,
\pl{388}{1996}{753}.}%

\lref\bjpsv{M. Bershadsky, A. Johansen, T. Pantev, V. Sadov and
C. Vafa, ``F-theory, Geometric Engineering and $N=1$ Dualities,''
hep-th/9612052.}%

\lref\newwitten{E. Witten, ``Solutions of Four-Dimensional Field
Theories Via M Theory,'' hep-th/9703166.}

\lref\swfour{N. Seiberg and E. Witten, ``Electric-Magnetic Duality, Monopole
Condensation and Confinement in $N=2$ Supersymmetric Yang-Mills
Theory,'' \np{426}{1994}{19}, hep-th/9407087 \semi
N. Seiberg and E. Witten, ``Monopoles, Duality and Chiral
Symmetry Breaking in $N=2$ Supersymmetric QCD,'' \np{431}{1994}{484},
hep-th/9408099.}%

\lref\egk{S. Elitzur, A. Giveon and D. Kutasov, ``Branes and $N=1$
Duality in String Theory,'' hep-th/9702014.}%

\lref\leeyi{K. Lee and P. Yi, ``Monopoles and Instantons in Partially
Compactified D-branes,'' hep-th/9702107.}%

\lref\bsv{M. Bershadsky, C. Vafa and V. Sadov, ``D strings on D
manifolds,'' \np{463}{1996}{398}, hep-th/9510225.}%

\lref\nonedual{N. Seiberg, ``Electric-Magnetic Duality in
Supersymmetric Nonabelian Gauge Theories,'' \np{435}{1995}{129},
hep-th/9611149.}%

\lref\bhoy{J. de Boer, K. Hori, Y. Oz and Z. Yin, ``Branes and Mirror
Symmetry in $N=2$ Supersymmetric Gauge Theories in Three Dimensions,''
hep-th/9702154.}%

\lref\ahiss{O.~Aharony, A.~Hanany, K.~Intriligator, N.~Seiberg and
M.~J.~Strassler,
``Aspects of $N=2$ Supersymmetric Gauge Theories in Three Dimensions,''
hep-th/9703110.}%

\lref\bdl{M. Berkooz, M. R. Douglas and R. G. Leigh, ``Branes
Intersecting at Angles,'' hep-th/9606139, \np{480}{1996}{265}.}%

\lref\ghm{M. Green, J. A. Harvey and G. Moore, ``I-brane Inflow and
Anomalous Couplings on D-branes,'' hep-th/9605053, \cqg{14}{1997}{47}.}%

\lref\bh{J. H. Brodie and A. Hanany, 
``Type IIA Superstrings, Chiral Symmetry, and $N=1$ 4D 
Gauge Theory Dualities,''
hep-th/9704043.}%

\lref\shimon{A. Brandhuber, J. Sonnenschein, S. Theisen, S. Yankielowicz,
``Brane Configurations and 4D Field Theory Dualities,''
hep-th/9704044.}%

\lref\IS{K. Intriligator and N. Seiberg,
``Mirror Symmetry in Three Dimensional Gauge Theories,''
hep-th/9607207, \pl{387}{1996}{513}.}%

\lref\asy{O. Aharony, J. Sonnenschein and S. Yankielowicz,
``Interactions of Strings and D-branes from M Theory,''
hep-th/9603009, \np{474}{1996}{309}.}%

\lref\fived{K. Intriligator, D. R. Morrison and N. Seiberg,
``Five-Dimensional Supersymmetric Gauge Theories and Degenerations of
Calabi-Yau Spaces,'' hep-th/9702198.}%

\lref\morsei{D. R. Morrison and N. Seiberg, ``Extremal Transitions and
Five-Dimensional Supersymmetric Field Theories,'' hep-th/9609070,
\np{483}{1996}{229}.}%

\lref\dkv{M. R. Douglas, S. Katz and C. Vafa, ``Small Instantons, del
Pezzo Surfaces and Type I' Theory,'' hep-th/9609071.}%

\lref\gms{O. J. Ganor, D. R. Morrison and N. Seiberg, ``Branes,
Calabi-Yau Spaces and Toroidal Compactification of the $N=1$ Six
Dimensional $E_8$ Theory,'' hep-th/9610251, \np{487}{1997}{93}.}%

\lref\oogvaf{H. Ooguri and C. Vafa, ``Geometry of $N=1$ Dualities in
Four Dimensions,'' hep-th/9702180.}%

\lref\ejs{N. Evans, C. V. Johnson and A. D. Shapere, ``Orientifolds,
Branes and Duality of 4D Gauge Theories,'' hep-th/9703210.}%

\lref\argdoug{P. C. Argyres and M. R. Douglas, ``New Phenomena in
$SU(3)$ Supersymmetric Gauge Theory,'' hep-th/9505062,
\np{448}{1995}{93}.}%

\lref\apsw{P. C. Argyres, M. R. Plesser, N. Seiberg and E. Witten,
``New $N=2$ Superconformal Field Theories in Four Dimensions,''
hep-th/9511154, \np{461}{1996}{71}.}%

\lref\aks{O. Aharony, S. Kachru and E. Silverstein, ``New $N=1$
Superconformal Field Theories in Four Dimensions from D-brane
Probes,'' hep-th/9610205,
\np{488}{1997}{159}.}%

\lref\barbon{J. L. F. Barbon, ``Rotated Branes and $N=1$ Duality,''
hep-th/9703051.}%

\lref\ads{I. Affleck, M. Dine and N. Seiberg, ``Dynamical
Supersymmetry Breaking in Supersymmetric QCD,'' \np{241}{1984}{493}\semi
I. Affleck, M. Dine and N. Seiberg,
``Dynamical Supersymmetry Breaking in Four Dimensions and its
Phenomenological Implications,'' \np{256}{1985}{557}.}%

\lref\sem{N. Seiberg, ``Electric-Magnetic Duality in Supersymmetric
Non-Abelian Gauge Theories,'' hep-th/9411149 , \np{435}{1995}{129}.}%

\lref\methree{O. Aharony, ``IR Duality in $d=3$ $N=2$ Supersymmetric
$USp(2N_c)$ and $U(N_c)$ Gauge Theories,'' hep-th/9703215.}%

\lref\karch{A. Karch, ``Seiberg duality in three dimensions,''
hep-th/9703172.}%
     
\lref\telaviv{A. Brandhuber, J. Sonnenschein, S. Theisen and
S. Yankielowicz, ``Brane Configurations and 4D Field Theory
Dualities,'' hep-th/9704044.}%

\lref\ahnoh{C. Ahn and K. Oh, ``Geometry, Branes and $N=1$ Duality in
Four Dimensions I,'' hep-th/9704061.}%

\lref\cggk{T. Chiang, B. R. Greene, M. Gross and Y. Kanter, ``Black
Hole Condensation and the Web of Calabi-Yau Manifolds,''
hep-th/9511204.}%

\lref\egkrs{S. Elitzur, A. Giveon, D. Kutasov, E. Rabinovici and
A. Schwimmer, ``Brane Dynamics and $N=1$ Supersymmetric Gauge
Theory,'' hep-th/9704104.}%

\Title{hep-th/9704170, RU-97-25, IASSNS-HEP-97/38}
{\vbox{\centerline{Branes, Superpotentials and Superconformal Fixed
Points}}}
\medskip
\centerline{Ofer Aharony$^1$ and Amihay Hanany$^2$} 
\vglue .5cm
\centerline{$^1$Department of Physics and Astronomy,}
\centerline{Rutgers University}
\centerline{Piscataway, NJ 08855-0849, USA}
\centerline{\tt oferah@physics.rutgers.edu}
\vglue .3cm
\centerline{$^2$School of Natural Sciences}
\centerline{Institute for Advanced Study}
\centerline{Princeton, NJ 08540, USA}
\centerline{\tt hanany@ias.edu}
\bigskip
\noindent

We analyze various brane configurations corresponding to field
theories in three, four and five dimensions.  We find brane
configurations which correspond to three dimensional $N=2$ and four
dimensional $N=1$ supersymmetric QCD theories with quartic
superpotentials, in which what appear to be ``hidden parameters'' play
an important role.  We discuss the construction of five dimensional
$N=1$ supersymmetric gauge theories and superconformal fixed points
using branes, which leads to new five dimensional $N=1$ superconformal
field theories.  The same five dimensional theories are also used, in
a surprising way, to describe new superconformal fixed points of three
dimensional $N=2$ supersymmetric theories, which have both
``electric'' and ``magnetic'' Coulomb branches.

\Date{4/97}  

\newsec{Introduction}

In the past year many interesting results, both in field theory and
in string theory, were discovered by studying the worldvolume dynamics
of branes in superstring theories. A particular construction was used
in \hw\ to study the dynamics of $N=4$ supersymmetric gauge theories
in $2+1$ dimensions. Variants of this construction have been applied in
\egk\ to $N=1$ theories in $3+1$ dimensions and in \newwitten\ to
solving a large class of $N=2$ theories in $3+1$ dimensions.

In this paper we analyze several new generalizations of the brane
construction of \hw, and discuss various problems which arise
in the translation
from the brane construction to the field theory. We begin in section 2
by analyzing 3D $N=2$ theories which are related by rotations of the
branes to the theories discussed in \hw. We find that the ``hidden
parameters'' discussed in \hw, corresponding to the $x_6$ positions of
the D-branes, are no longer hidden, but affect the superpotential in
some configurations. The precise form of this effect is not clear, but
we provide consistent superpotentials for describing the low-energy
theory of general brane
configurations. We discuss also the generalization of these results to
4D $N=1$ theories, and their consistency with the brane constructions
of mirror symmetry \refs{\IS,\hw}, Seiberg duality \refs{\sem,\egk}
and chiral symmetry \bh.

In section 3, which is independent of section 2, we discuss the
generalization of the construction of \hw\ to five dimensional field
theories with $N=1$ supersymmetry. We find many interesting effects,
related to the recent analysis of 5D $N=1$ field theories in
\refs{\seibergfive,\morsei,\dkv,\fived}, and provide (implicit) 
constructions for new 5D $N=1$ superconformal field theories.

In section 4 we use the brane constructions of section 3 with
additional D3-branes to construct new 3D $N=2$ superconformal field
theories, which have both ``electric'' and ``magnetic'' Coulomb
branches emanating from a single point in their moduli space. Every 5D
superconformal theory we construct in section 3 gives rise to a 3D
$N=2$ superconformal theory, opening a way
to a systematic study of
a large class of new three dimensional theories.

\newsec{Branes and Superpotentials in 3D $N=2$ Supersymmetric Gauge Theories}

\subsec{The brane construction}
\subseclab{\brane}

In \hw\ a configuration of 3-branes and 5-branes was used to construct
$N=4$ supersymmetric gauge theories in three dimensions. In this
section we will generalize this brane configuration to a construction
of $N=2$ gauge theories in three dimensions. First, we recall the
configuration of \hw.  The configuration is in the type IIB string
theory with time coordinate $x_0$ and space coordinates $x_1,\ldots,
x_9$. We denote by $Q_L$ and $Q_R$ the supercharges generated by left-
and right- moving world-sheet degrees of freedom. The type IIB theory
is chiral, and the supercharges obey $\bar\Gamma Q_L=Q_L$ and
$\bar\Gamma Q_R=Q_R$ with $\bar\Gamma=\Gamma_0\cdots \Gamma_9$.  The
brane configuration of \hw\ includes NS 5-branes which span a
worldvolume in the 012345 directions, D5-branes which span a
worldvolume in the 012789 directions and 3-branes which span a
worldvolume in the 0126 directions.  This configuration preserves 1/4
of the supersymmetry. The 3-branes are finite in one direction and
thus their worldvolume theory leads to $N=4$ supersymmetry in three
dimensions.  Breaking the supersymmetry by a further half may be
achieved by introducing a rotation angle \bdl\ to at least one of the
branes. We can choose the rotation to be in the 45 - 89 space,
i.e. rotate by an angle $\theta$ in the $x_4-x_8$ and $x_5-x_9$
planes.  In fact, we can rotate each of the 5-branes by different
rotation angles without breaking further the supersymmetry (below 3D
$N=2$)\foot{This was also discussed in
\barbon.}. We will discuss general situations of this sort.
In a particular case where we
choose the angles of the NS 5-branes to be 0 and 90 degrees while
the angles of the D5-branes are zero (relative to the configuration
of \hw\ described above) the configuration is T-dual
(in the $x_3$ direction) to the configuration described in \egk. This
configuration was also studied in \bhoy.

In the special cases where the NS 5-branes have zero angle,
i.e. with worldvolume coordinates spanning the 012345 directions, we
will call the 5-brane a NS brane, while for 90 degrees,
i.e. worldvolume 
coordinates spanning the 012389 directions, we will call the
5-brane a NS$'$ brane.  Similarly, when the D5-branes have zero angle
with worldvolume coordinates 012789 we will 
call them D 5-branes, and for
90 degrees with worldvolume coordinates 012457 we will call them D$'$
5-branes.

The unbroken supersymmetry generators are linear
combinations $\epsilon_L Q_L + \epsilon_R Q_R$ which satisfy
\eqn\susygennsone{\eqalign{
\epsilon_L & = \Gamma_0 \Gamma_1 \Gamma_2
\Gamma_3 \Gamma_4 \Gamma_5 \epsilon_L \cr
\epsilon_R & = -\Gamma_0 \Gamma_1 \Gamma_2
\Gamma_3 \Gamma_4 \Gamma_5 \epsilon_R \cr
}}
due to the presence of a NS brane,
\eqn\susygennstwo{\eqalign{
\epsilon_L & = \Gamma_0 \Gamma_1 \Gamma_2
\Gamma_3 \Gamma_8 \Gamma_9 \epsilon_L \cr
\epsilon_R & = -\Gamma_0 \Gamma_1 \Gamma_2
\Gamma_3 \Gamma_8 \Gamma_9 \epsilon_R \cr
}}
due to the presence of a NS$'$ brane,
\eqn\susygendfiveone{\eqalign{
\epsilon_L &= \Gamma_0 \Gamma_1 \Gamma_2 \Gamma_7 \Gamma_8 \Gamma_9 \epsilon_R
\cr}}
due to the presence of a D 5-brane,
\eqn\susygendfivetwo{\eqalign{
\epsilon_L &= \Gamma_0 \Gamma_1 \Gamma_2 \Gamma_4 \Gamma_5 \Gamma_7 \epsilon_R
\cr}}
due to the presence of a D$'$ 5-brane, and
\eqn\susygendone{\eqalign{
\epsilon_L &= \Gamma_0 \Gamma_1 \Gamma_2 \Gamma_6 \epsilon_R
\cr}}
due to the presence of the D3-branes. Additional branes can also be
added without breaking the supersymmetry any further, but we will not
discuss them here.

\fig{Three dimensional $N=2$ supersymmetric gauge theories with gauge
group $U(N_c)$ and $N_f$ quarks.  There are $N_c$ D3-branes
(horizontal lines), which are stretched in between two NS 5-branes
(vertical lines). The figure is depicted in the 36 plane as indicated
by the arrows in the upper right of the figure.  The left 5-brane
stretches along the 012345 directions and is denoted NS and the right
5-brane stretches along the 012389 directions and is denoted NS$'$.
The ``X''s denote D5-branes, and the ``+''s denote D$'$ 5-branes, both
of which give rise to quarks.}
{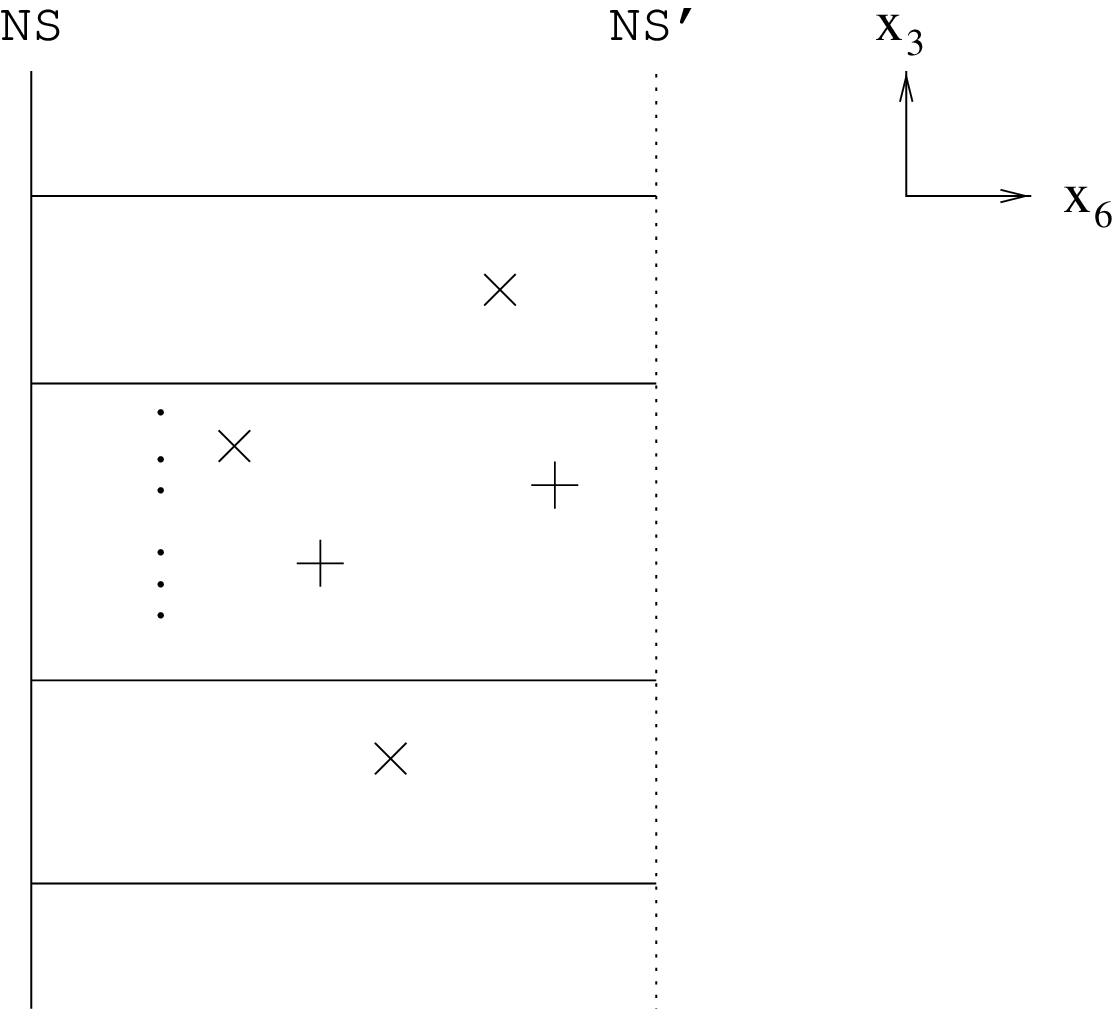}{10 truecm}
\figlabel\YM

The presence of all these branes breaks the Lorentz group $SO(1,9)$ to
$SO(1,2)\times SO(2)_{45} \times SO(2)_{89}$, where $SO(1,2)$ acts on
$x_0,x_1$ and $x_2$, $SO(2)_{45}$
acts as rotations in the $x_4,x_5$ plane and $SO(2)_{89}$ acts as rotations in
the $x_8,x_9$ plane.
The $SO(2)$ symmetries are broken when some of the branes are at angles
different from 0 or 90 degrees.

As in \hw\ we introduce the mirror symmetry operation. The $S$
transformation in the $SL(2,\IZ)$ U-duality group of the type IIB
string theory, given by the matrix
\eqn\smatrix{\pmatrix{0&1\cr-1&0\cr},} 
together with the rotation which maps
$x_i$ to $x_{i+4}$ and $x_{i+4}$ to $-x_i$ ($i=3,4,5$), define an
operation which we call mirror symmetry.

\subsec{The field content of the 3D theory}

As in \hw\ we will take the D3-branes to be finite in one
direction. Their low energy theory (below the scale determined by the
distance between the NS 5-branes) is then described by a three
dimensional field theory.  One might expect this theory to be the
dimensional reduction of the 10D SYM theory, which gives $N=8$ SYM in
three dimensions. However, as in \hw, boundary conditions on some of
the fields where the D3-brane ends on the 5-branes remove these
fields from the low energy theory.

The boundary condition for a D3-brane ending on a NS(012345)
5-brane projects out the motions of the D3-brane in the $x_7,x_8$ and
$x_9$ directions, as well as the $A_6$ component of the D3-brane gauge
field. The 789 components are set to the values of the position of the NS
brane in these directions, and $A_6$ is also frozen (but it is not
clear at which value).
The boundary conditions coming from a NS(012389) 5-brane
project out the $x_4,x_5$ and $x_7$ scalars in the field theory, as
well as the $A_6$ component of the gauge field.
The 457 components are set to the values of the position of the NS
brane in these directions.
In order to generalize these results to NS 5-branes at arbitrary
angles, we introduce complex coordinates $z_1=x_4+ix_5$ and
$z_2=x_8+ix_9$.  The 3-brane worldvolume theory will include scalar
fields $\phi_1$ and $\phi_2$, respectively, which denote the position
of the 3-brane in these directions. These are the scalar components of
chiral multiplets in the 3D $N=2$ theory.  Two more scalar
parameters in the D3-brane worldvolume theory 
are the $x_7$ coordinate and the $A_6$ component of
the gauge field.  To understand what kind of multiplet these scalars
form we recall that mirror symmetry exchanges the scalars $\phi_1$ and
$\phi_2$.  Mirror symmetry also acts as electric-magnetic duality on
the worldvolume of the 3-brane. Thus, it exchanges the $x_7$
coordinate of the 3-brane and the $x_6$ component of the gauge
field with the $x_3$ coordinate of the 3-brane and the scalar dual to
the 3D gauge field, respectively.
The scalars $x_7$ and $A_6$ are dual to
a vector multiplet which includes $A_\mu$ and $x_3$. 
To some extent the scalar
$x_7+iA_6$ can be treated as the lowest component of a chiral
multiplet $\phi_3$.  There is a $U(1)$ symmetry which acts by shifting
the $A_6$ component of the gauge field (which is periodic in the
quantum theory).

The boundary conditions for a D3-brane ending on a NS 5-brane set
$\phi_3$ and a linear combination of $\phi_1$ and $\phi_2$ (which
depends on the angle of the NS 5-brane) to zero at the boundary. The
boundary conditions for a D3-brane ending on a D5-brane likewise set
to zero the vector multiplet and a linear combination of $\phi_1$ and
$\phi_2$ depending on the angle of the D5-brane. From
the three dimensional point of view, each of these fields will give
rise to a tower of massive states corresponding to the momenta in the
$x_6$ direction, the lowest of which will have a mass of order
$1/|\Delta x_6|$ where $\Delta x_6$ is the span of the D3-brane in the
$x_6$ direction. The relation of $\Delta x_6$ to the gauge coupling
\hw\ sets this mass to be of the order of $g^2$, where $g$ is the gauge
coupling. In \hw\ (and in other cases discussed in the literature),
the presence of these massive states was not important for the
low-energy field theory. However, as we will see in \S2.4, in some cases
these states will give rise to important interactions in the
low-energy field theory, and they cannot be ignored.

Now, we can consider which fields remain massless in the D3-brane
worldvolume theory, beginning with the vector multiplet and $\phi_3$.
As in \hw\ there are three cases to consider :
\item{1)} When a 3-brane has both ends on NS 5-branes, the $N=2$ vector
multiplet survives while the chiral multiplet $\phi_3$ is frozen 
(i.e. massive).
In this case the $x_7$ coordinates of the 5-branes must coincide to
preserve supersymmetry. The $x_6$ distance between the two NS 5-branes
is identified with $1/g^2$ (up to the string coupling), and the $x_7$
distance between them is identified with the \FI\ term for the gauge
field.
\item {2)} When a 3-brane has both ends on D5-branes, the $N=2$ chiral
multiplet $\phi_3$ survives while the vector multiplet is frozen.
In this case the $x_3$ coordinates of the D5-branes must coincide to
preserve supersymmetry.
\item{3)} When a 3-brane ends on a D5-brane at one side and on a NS
5-brane at the other side both $\phi_3$ and the vector multiplet are frozen.

We need also to see what are the remaining massless fields
which are associated with the positions along the 45 and 89 directions.
For this the analysis does not distinguish between the various cases
described above.
Given a 3-brane with both ends on 5-branes whose 45-89 position
is given by an
equation $az_1+bz_2=\alpha$ for the first 5-brane, and an
equation $cz_1+dz_2=\beta$ for the other 5-brane,
there are several cases to consider :
\item{i)} If $ad-bc\not=0$ there is a unique solution to the two
equations, and both scalar fields are frozen to the values of the 
solution. There are no additional moduli beyond the ones described above.
\item{ii)} If $ad-bc=0$ with a
one parameter solution to the two equations,
there is one massless chiral multiplet corresponding to the motion of
the D3-brane in the direction perpendicular to $az_1+bz_2$.
This chiral multiplet combines with the other multiplets 
considered in cases 1) to 3) above
to form :
\item{1)} an $N=4$ vector multiplet;
\item{2)} an $N=4$ hypermultiplet;
\item{3)} an $N=2$ chiral multiplet.
\item{iii)} If $ad-bc=0$ with no solution to the two equations, there is no
supersymmetric configuration of a D3-brane between the two 5-branes.
Such a situation corresponds, as in \hw, to turning on 
\FI\ parameters for the
$N=4$ gauge theory (which look like a superpotential proportional to
$b\phi_1-a\phi_2$ in the $N=2$ theory), leading to supersymmetry breaking.

Thus, we have three cases in which : \item{i)} There are no scalar
moduli from $\phi_1$ and $\phi_2$;
\item{ii)} There is one complex modulus; \item{iii)} Supersymmetry is broken.

The values of the angles are generally
not relevant parameters for the low energy field theory,
since if we change the angles the massless content does not
change, except in particular cases where branes become parallel.
The angles do, however, affect the massive modes of the field theory.
In the discussion below we will specialize in most cases 
to angles of either 0 or 90 degrees.

As in \hw, it is interesting to examine what a configuration of
D3-branes stretched between a NS 5-brane and a NS$'$ 5-brane looks
like from the 5-brane point of view. Recall that in \hw, a D3-brane
ending on a 5-brane looked like a magnetic monopole. So, if we have
$N_c$ D3-branes stretched between the two 5-branes, from the
point of view of each 5-brane we have a configuration of magnetic
charge $N_c$ under the $U(1)$ gauge field living on the
5-brane. However, in this case, unlike the case described in \hw,
bringing the NS 5-branes together in the 67 directions (in the
field theory this means taking 
the gauge coupling to infinity and the \FI\ term to zero)
does not correspond to an enhanced
$U(2)$ gauge symmetry, but instead to a theory which has
charged matter under the two $U(1)$ groups living on the NS 5-branes.
This charged matter arises from D-strings, in the same way that for
intersecting D-branes we get charged matter from fundamental strings.
It consists of one hypermultiplet with charges $(1,-1)$ and $(-1,1)$ under
$U(1)\times U(1)$.
The extra scalars may serve as additional parameters for the strongly
coupled 3D field theory.
There is no simple interpretation of the moduli space of this
configuration as a monopole moduli space of some gauge theory, as in
\hw.

\subsec{Global symmetry}

As mentioned in \S\brane, two global symmetries that are apparent in the
construction when all angles are chosen to be zero or 90 degrees
are $SO(2)_{45}$ and $SO(2)_{89}$. The scalar components
of the fields $\phi_1$ and $\phi_2$ described above obviously have
charges $(2,0)$ and $(0,2)$ under these symmetries (in a convenient
normalization), while the other bosonic fields arising from D3-D3
strings are uncharged. The fermionic fields arising from D3-D3 strings
are the reduction of a 10D spinor, so they have charges $+1$ or $-1$
under both $U(1)$ symmetries. We can identify $SO(2)_{89}$ with the
$U(1)_R$ symmetry of the 3D $N=2$ theory, and then $SO(2)_{45}$ is a
combination of this $U(1)_R$ symmetry with a global symmetry under which
$\phi_1$ and $\phi_2$ are oppositely charged.

When D5-branes are also present, quarks
arise from open strings between the D3-branes and the D5-branes. For a
D5-brane in the 012789 directions, these strings
have Dirichlet-Neumann (DN) boundary conditions in the $x_6,x_7,x_8$
and $x_9$ directions, and DD or NN boundary conditions in all other
directions. Thus, the states from the NS sector of the open strings,
which we identify as the bosons in the quark multiplets, will be
spinors of $SO(4)_{6789}$, and they will be charged, in particular,
under $SO(2)_{89}$, but not under $SO(2)_{45}$. The fermions, on the
other hand, come from the R sector of the open strings, and they are
uncharged under $SO(4)_{6789}$ (and, therefore, also under
$SO(2)_{89}$), but they will be charged under $SO(2)_{45}$ (since they
arise, in the RNS formalism, from the quantization of RR zero modes in
these directions). For
D5-branes in the 012457 directions (D$'$ branes), 
$SO(2)_{45}$ and $SO(2)_{89}$ will
be interchanged in this discussion.

In the construction of \hw, there was an additional $U(2)$ global
symmetry associated with coincident NS 5-branes (which was just the gauge
symmetry of these 5-branes), that was broken to $U(1)^2$ for finite
gauge coupling, but this does not appear in our case, as
described at the end of the last section.

\subsec{$U(1)$ gauge theory with $N_f$ flavors}

We begin our analysis of 3D $N=2$ gauge theories by considering a
$U(1)$ theory with $N_f$ flavors. The $U(1)$ gauge theory arises by
considering a D3-brane stretched between two NS 5-branes, which are
generally described by some equation $a_j z_1 + b_j z_2 = c_j$
($j=1,2$) in the $4589$ complex plane.  The quark flavors arise from
$N_f$ D5-branes, which are similarly described by parameters $\tilde
a_i, \tilde b_i,\tilde c_i$ ($i=1,\cdots,N_f$), and by additional
parameters $z_i$ and $\tilde{m}_i$ corresponding to their $x_6$ and
$x_3$ positions (we will assume $\tilde{m}_i=0$ in the discussion
below).

We would like to write down a superpotential describing this theory,
including the quarks and the massless and massive
$\phi_i$ fields. One term that obviously exists in such a
superpotential is the coupling of the quarks to the adjoint fields, of
the form
\eqn\UoneNf{W=\sum_{i=1}^{N_f}(\tilde a_i\phi_1+\tilde b_i\phi_2-\tilde c_i)
\tilde Q^i Q_i,}
which is determined by the local $N=4$ supersymmetry of the D3-D5
system (we normalize $\tilde{a}_i^2 + \tilde{b}_i^2 = 1$); the
$\tilde{c}_i$ parameters have the obvious interpretation as quark
masses, and we will set them (as well as the $c_j$) to zero in the
discussion below (they can always be added later).  However, it is
clear that it is not always possible to include all the interactions
in a superpotential. For instance, in the $N=4$ theory of \hw, the
boundary conditions give a mass to $\phi_2$ and $\phi_3$, but it is
impossible to write down a superpotential that incorporates this in an
$N=4$ invariant way (or even in a way which is invariant under the
global symmetries, including the shift symmetry of the complex part of
$\phi_3$). This is not too surprising since $\phi_3$ is in some sense
actually a dual vector multiplet, so we would not expect to be able to
write all its interactions in a superpotential. In any case, it is
clear that in the configuration described above, the boundary
condition on one side makes (in some way) $a_1 \phi_1 + b_1 \phi_2$
and $\phi_3$ massive, while the boundary condition on the other side
makes $a_2 \phi_1 + b_2 \phi_2$ and $\phi_3$ massive. Since we do not
know how to write down these mass terms explicitly, we cannot
explicitly integrate out the massive fields and write down the low
energy effective theory. However, we will assume that some consistent
procedure for doing this exists, which will naturally give rise
(starting from \UoneNf\ and mass terms for the $\phi$'s) to quartic
interactions between the quarks, and we will determine the low-energy
superpotentials to fit the global symmetries and the moduli spaces we
find in the various cases.

To simplify the discussion, we will assume that the two NS 5-branes are
at zero and 90 degrees, i.e. they are a NS and NS$'$ brane (the
generalization of our results to other cases should be
straightforward). In this case the first boundary condition gives a
mass to $\phi_2$ and $\phi_3$, and the second to $\phi_1$ and
$\phi_3$. While we cannot describe these mass terms by a
superpotential, a superpotential of the form $\mu \phi_1 \phi_2$ is
consistent with the global symmetries described above, so we will
assume that such a term is formed from the ``boundary mass terms''
when integrating out $\phi_3$. This assumption will pass several
consistency checks, as described below. Since we expect the masses of
$\phi_1$ and $\phi_2$ to be of order $g^2$, we expect $\mu$ also to be
of this order\foot{Note that in \bh\ it was assumed that the masses of
the $\phi$'s were proportional to $\tan(\theta)$, where $\theta$ was
the angle between the two NS 5-branes, and this passed some
consistency checks. We expect this to be true only for small angles,
when the mass is much smaller than the scale $\mu$ discussed here.}. 
Of course, there is really an infinite tower of states
corresponding to $\phi_1$ and $\phi_2$ arising in the dimensional
reduction from the 4D theory to the 3D theory, but it turns out that
(at least for our current purposes) using the naive mass term $\mu
\phi_1 \phi_2$ will suffice.

Let us begin with the simplest case of $N_f=1$ (the $N_f=0$ theory is
obviously trivial, with just one massless vector multiplet). Adding
the superpotentials we described above, we find for this theory
\eqn\Uoneone{W = \mu \phi_1 \phi_2 + (\tilde{a}_1 \phi_1 + \tilde{b}_1
\phi_2) Q \tilde{Q}.}
Now, we can integrate out $\phi_1$ and $\phi_2$. If $\tilde{a}_1$ or
$\tilde{b}_1$ are zero, so that the D-brane is oriented at zero or 90
degrees, this leaves no superpotential in the low-energy theory, which is just
the $U(1)$ $N_f=1$ gauge theory discussed in \ahiss, which has a one
dimensional Higgs branch (as well as Coulomb branches). However, at
different angles, we find a low-energy superpotential of the form $W
\sim {\tilde{a}_1\tilde{b}_1\over \mu} (Q \tilde{Q})^2$, which lifts the Higgs
branch. Note that in these cases the $SO(2)_{45}$ and $SO(2)_{89}$ symmetries
are explicitly broken, so this superpotential is consistent with the
global symmetries. The parameter in front of the quartic term is
dimensionless, corresponding to $g^2/\mu$ if we had kept the
dependence of the superpotential on the gauge coupling, and we expect
it to be of order one for a generic angle. 
Comparing these results to what we see in the
brane picture, we find an exact agreement.  Only when the D5-brane is
parallel to either the NS brane or the NS$'$ brane (in the 45-89 plane)
is there a flat direction corresponding to splitting the D3-brane at
the D5-brane, which we identify with the Higgs branch of the field
theory. Note that, in the full theory including the massive fields, 
the field parametrizing
this branch is really not just $M = Q \tilde{Q}$, but in fact a
combination of this field and the chiral superfield corresponding to
the motion of the D3-brane in the direction it moves along in the
Higgs branch. This will be the case in all the examples described
here.

Things become slightly more complicated when additional D-branes are
added. For simplicity we will assume from here on that all D-branes
are also oriented at zero or 90 degrees. This is the case where the
Higgs branches have the maximal dimension, and the generalization to
arbitrary angles should again be straightforward. For $N_f=2$ there
are now four different possibilities, depending on the orientation of
the branes -- in an obvious notation, corresponding to the ordering of
the branes in $x_6$, they are NS-D-D-NS$'$,
NS-D-D$'$-NS$'$, NS-D$'$-D-NS$'$ and NS-D$'$-D$'$-NS$'$. In the first case, the
naive superpotential is 
\eqn\Uonetwopar{W = \mu \phi_1 \phi_2 + \phi_1 (Q_1
\tilde{Q}^1 + Q_2 \tilde{Q}^2),}
and integrating out the massive fields leaves no low-energy
superpotential. This is consistent with the brane moduli space in this
case, as discussed in \refs{\bhoy,\ahiss}. The fourth case is related
to this one by an obvious rotation and reflection, and behaves in the
same way. In the second and third cases, if we choose the first quark
to come from the D brane and the second quark to come from the
D$'$ brane, we would write a superpotential
\eqn\Uonetwonotpar{W = \mu \phi_1 \phi_2 + \phi_1 Q_1 \tilde{Q}^1 +
\phi_2 Q_2 \tilde{Q}^2}
to describe both cases. Integrating out $\phi_1$ and $\phi_2$, we find
$W \sim {1\over \mu} Q_1 \tilde{Q}^1 Q_2 \tilde{Q}^2$, and we can
analyze the field theory with this superpotential by the methods used
in \ahiss. We find that the moduli space of this theory consists of 6
one dimensional branches intersecting at the origin of moduli space,
each of which is parametrized by one of $V_+, V_-, M^1_1, M^1_2,
M^2_1$ and $M^2_2$ (where $M^j_i = Q_i \tilde{Q}^j$). 
This moduli space is actually the same (up to the
quantum splitting of the Coulomb branch which is not visible in the
brane picture) as that of the NS-D-D$'$-NS$'$ configuration. The Coulomb
branch corresponds to a D3-brane connecting the NS and NS$'$ branes
(free to
move in the $x_3$ direction), the $M^1_1$ branch corresponds to
splitting the D3-brane at the D brane (and then the D-NS$'$ piece of it
is free to move in $x_{89}$), the $M^2_2$ branch corresponds to
splitting the D3-brane at the D$'$ brane (and then the NS-D$'$ piece of it
is free to move in $x_{45}$), while the $M_1^2$ and $M_2^1$ branches
correspond to splitting the D3-brane twice, so that its middle part is
free to move in $x_7$. Naively the double splitting corresponds to a
single branch, but since this branch is identified with the Coulomb
branch of the mirror theory \refs{\IS,\hw,\bhoy,\ahiss} it is clear that
it should be split in the quantum theory corresponding to this brane
configuration (as described in
\ahiss). Therefore, the NS-D-D$'$-NS$'$ configuration is consistently
described by the superpotential \Uonetwonotpar. However, this
superpotential does not consistently describe the NS-D$'$-D-NS$'$
configuration, in which a double splitting of the D3-brane leads to a
three dimensional branch. In fact, this configuration is consistently
described by the theory with no superpotential. Note that the only
difference between these configurations is in the relative $x_6$
positions of the D-branes, which were called ``hidden parameters'' in
\hw\ (and did not affect the low-energy field theory there). In the
configurations of the type we discuss, it is clear that these
parameters (or at least their signs) do affect the low-energy field
theory, since they change the Higgs branches as discussed above. Since
we do not know how to include the effects of these hidden parameters
on the low-energy theory, we will again have to be content with
guessing the results. In the $N_f=2$ case, a consistent assumption
seems to be that the quartic superpotential is generated for one sign
of the $x_6$ distance between the D and D$'$ 5-branes, corresponding to
the NS-D-D$'$-NS$'$ configuration, but no superpotential is generated for
the other sign.

The results for other values of $N_f$ are similar to the $N_f=2$ case,
and depend again on the ordering of the D and D$'$ branes. When all
D5-branes have the same orientation, no superpotential is generated
(no superpotential is consistent with the global symmetries in this
case), and this is consistent with the moduli space of the brane
construction. When some of the branes are D (leading to quarks
$Q_i,\tilde{Q}^i$) and some are D$'$ (leading to quarks $q_j,
\tilde{q}^j$), superpotentials of the form $W \sim Q_i \tilde{Q}^i
q_j \tilde{q}^j$ may be generated as above (they are consistent with
the global symmetries). For instance, the NS-D$^n$-(D$'$)$^m$-NS$'$
configuration is consistently described by a superpotential of the
form $W \sim (\sum_{i=1}^n Q_i \tilde{Q}^i) (\sum_{j=1}^m q_j
\tilde{q}^j)$, while the NS-(D$'$)$^m$-D$^n$-NS$'$ configuration is
consistently described by $W=0$. Note that these superpotentials are
consistent with the $SU(n)$ and $SU(m)$ global symmetries that appear
in the brane picture when we take all the D branes and all the
D$'$ branes together (the chiral global symmetries of these
configurations will be discussed in \S2.8). In other orderings of
the branes, we can no longer take the D branes and D$'$ branes
together without interchanging them (which is a phase transition we do
not know how to analyze), and the superpotentials no longer respect
these symmetries. Since D3-branes between two D branes or between two
D$'$ branes have two chiral superfields describing their position,
while D3-branes between a D brane and a D$'$ brane have only one, it
is clear that the maximum dimension of the Higgs branch is obtained in
the NS-(D$'$)$^m$-D$^n$-NS$'$ configuration, while smaller Higgs
branches arise when the D and D$'$ branes are interposed. For each
brane configuration we can write down some quartic superpotential that
gives the correct moduli space, but the rules for doing so are not
clear since we do not know the effect of the ``hidden'' $x_6$
parameters on the low-energy field theory.

The dependence of the superpotential on the $x_6$ ordering of the
branes is consistent with the fact that the branes intersect each
other when we move them past each other in the $x_6$ direction, so a
non-trivial phase transition may occur. However, we can also take the
D-branes around each other by moving them in the $x_3$ direction,
which corresponds to giving a real mass to the quark before exchanging
the branes. Obviously, once the quarks have a real mass (relative to
the position of the D3-brane), they cannot obtain VEVs, so the
superpotentials we write are no longer meaningful in this
case. However, there should still be some continuous interpolation
between the different superpotentials we write upon adding (and then
removing) real masses for the quarks and shifting the branes in
$x_6$. Similar problems arise when
exchanging D and NS$'$ branes in the $x_6$ direction (also there a phase
transition may naively be avoided after giving the quarks a
mass), and they will be discussed in \S2.9.

\subsec{$U(N_c)$ gauge theories}

The discussion of $U(N_c)$ theories with general angles for the
D-branes is similar to the previous discussion, but with some
important differences. Generally, we expect to get also in these
theories quartic superpotentials for the quarks, and we expect that 
after taking these into account
the field theory moduli spaces will agree with the
brane construction. However, for $N_c > 1$, the global symmetries no
longer uniquely determine the form of the superpotentials, even in the
simplest cases.

An important difference between the $U(1)$ case and the non-Abelian 
cases is
the existence of Euclidean-string-instanton configurations, which can
lift part of the moduli space. For example, when two D3-branes are
stretched between a NS and a NS$'$ 5-brane with no D5-branes between
them, a superpotential is generated that drives these branes
apart\foot{The boundaries of the D3-branes in the 5-branes look like
4D $N=2$ monopoles, but the presence of the other 5-branes breaks the
$N=2$ to $N=1$, so there is no longer any reason for the forces between
these monopoles to cancel.}, corresponding to the
Affleck-Harvey-Witten instanton-generated superpotential in field
theory
\refs{\ahw,\bhoy,\ahiss}. In this case the instanton effects in the
brane picture agree with our low-energy field theory
description. However, another case where a superpotential appears to
be generated in the brane picture is when D3-branes are stretched
between D and D$'$ 5-branes. Since this is the $SL(2,\IZ)$ dual of the
previous case, we expect Euclidean fundamental strings to generate a
superpotential in this case that will drive the D3-branes apart and
lift the corresponding Higgs branches. From the point of view of the
mirror theory, this superpotential is the standard
Affleck-Harvey-Witten superpotential, but in the original theory it is
an instanton effect lifting part of the Higgs branch, which we do not
know how to describe in the low-energy theory (it depends on the
``magnetic gauge coupling'' in the language of \hw, which we do not
know how to identify in the conventional field theory description).
Surprisingly, the low-energy field theory superpotentials are
consistent with simply ignoring those branches of the brane picture
which are apparently lifted by such instanton effects, as we describe
below.

For example, let us analyze in detail the $U(2)$ theory with
$N_f=2$. If the two D-branes are parallel, the arguments above (and
the global symmetries) suggest that there is no superpotential, and
the brane moduli space is consistent with this assumption. In the
NS-D-D$'$-NS$'$ and NS-D$'$-D-NS$'$ configurations, we again seem to
have a superpotential of the form
\eqn\Utwotwonotpar{W = \mu \tr(\phi_1 \phi_2) + Q_1 \phi_1 \tilde{Q}^1 +
Q_2 \phi_2 \tilde{Q}^2,}
where $\phi_1$ and $\phi_2$ are now $U(2)$-adjoint
matrices. Note that the D3-brane worldvolume theory now includes also
an interaction of the form $W \sim 
\tr([\phi_1,\phi_2]\phi_3)$, whose effect
on the low-energy dynamics is not clear. In any case, integrating out
$\phi_1$ and $\phi_2$ in \Utwotwonotpar, we find a superpotential of
the form $W \sim M^2_1 M^1_2$. Unlike
the $U(1)$ case, however, this is not the most general superpotential
consistent with the global symmetries,
and a term of the form $W \sim M^1_1 M^2_2$ might also appear. 

\fig{Some Higgs branches of $U(2)$ theories with two flavors, in a
schematic representation (the horizontal axis is $x_6$, but the
vertical axis corresponds to no particular coordinate)
: (a) The most general Higgs branch of the NS-D-D$'$-NS$'$ configuration,
(b) A Higgs branch lifted by instantons in the NS-D$'$-D-NS$'$
configuration, (c) An unlifted Higgs branch in the NS-D$'$-D-NS$'$
configuration.}{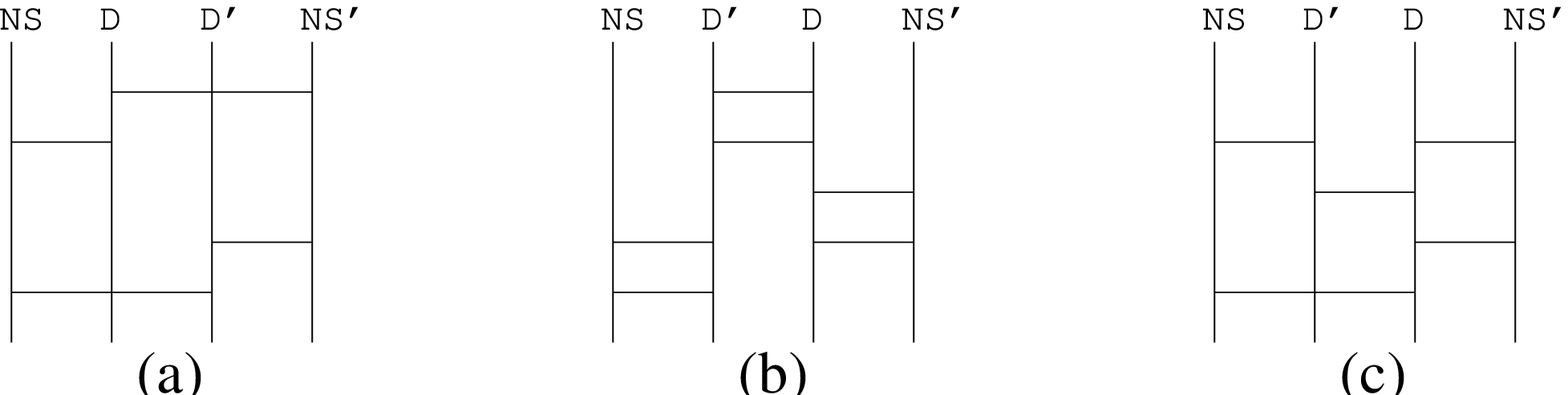}{15 truecm}
\figlabel\Higgs

As we found for the $U(1)$ case, the moduli space of the
NS-D-D$'$-NS$'$ configuration is consistent with the presence of
superpotentials of this form.  Using the fact (required for
consistency \hw) that there are no supersymmetric $s$-configurations
(in which more than one D3-brane is stretched between the same D and
NS branes, or between the same D$'$ and NS$'$ branes),
the most
general Higgs branch of this configuration, which is two dimensional,
appears in figure
\Higgs(a). We can identify this branch 
(assuming the superpotential is just $W
\sim M^2_1 M^1_2$) with the Higgs branch of the field theory, where
$M^1_1$ and $M^2_2$ are non-zero. We can also analyze the Coulomb
branches and the mixed branches, by adding the superpotential which
consistently describes the quantum moduli space of this theory, of the
form $W = V_+ V_- (M^1_1 M^2_2 - M^1_2 M^2_1)$ \ahiss. For instance,
the field theory has a branch where $M^1_1$ and $V_-$ are both
non-zero. In the brane picture, this corresponds to splitting one of
the 3-branes twice, and leaving the other 3-brane continuous (as
before, a quantum splitting which is not visible in the naive brane
picture means that the same type of configuration, with different
brane positions, describes also the branch where $M^2_2$ and $V_-$ are
non-zero).

The analysis of the NS-D$'$-D-NS$'$ configuration is more complicated
here, since some of its branches, like the one depicted in figure
\Higgs(b), which is naively six dimensional, 
are lifted by instanton effects. However, the remaining
branches seem to correspond to those of the field theory with $W=0$,
as we found also in the $U(1)$ case. For example, the unlifted Higgs
branch depicted in figure \Higgs(c) is four dimensional, which is the
same dimension we find in the field theory. Higgs branches with 
an unbroken $U(1)$ are three dimensional in the brane
configuration as well as in the field theory.

Similar considerations apply also for higher $N_c$ and $N_f$. Consider,
for instance, higher values of $N_c$ with $N_f=2$. For $N_c > 3$ there
are no unlifted branches in the NS-D$'$-D-NS$'$ configuration,
consistent with the assumption of $W=0$ in this case, since the
corresponding field theory has no supersymmetric vacua either
\ahiss. For $N_c=3$ there are unlifted five dimensional Higgs-Coulomb
branches (with D3-brane segments connecting, for instance, NS-NS$'$,
NS-D$'$, D$'$-D, D-NS$'$, NS-D$'$ and D$'$-NS$'$), in agreement with
the field theory constraint $V_+ V_- \det(M) = 1$
\ahiss. In the NS-D-D$'$-NS$'$ configuration of $U(3)$ with $N_f=2$ we
find only a three dimensional Higgs-Coulomb branch, in agreement with
the constraint and the additional superpotential $W \sim M^2_1 M^1_2$
which sets $M^2_1 = M^1_2 = 0$. For higher values of $N_f$, as in the
$U(1)$ case, more general superpotentials arise, depending on the
ordering of the branes. For each ordering there seems to be a
particular superpotential, consistent with the global symmetries, that
correctly gives the brane moduli space.

\subsec{Mirror symmetry}

The configurations we discuss here can be used to construct mirror
symmetries of three dimensional gauge theories \IS, in the same way as
in \hw\ for $N=4$ theories and in \refs{\bhoy,\ahiss} for 
$N=2$ theories. However, we can also derive these ``new'' mirror
symmetries directly from the known mirror symmetries of the $N=2$ SQCD
theory. Since the theories we discuss differ from the standard SQCD
theories just by having an additional superpotential, their mirror is
just the mirror of the standard SQCD theory, plus a superpotential
which is the image of the additional superpotential under the
mirror transformation. This is analogous to the relation between the
$N=4$ mirror symmetry and the $N=2$ mirror symmetry \ahiss.

For example, let us look at the $U(1)$ theory with $N_f=2$ in the
NS-D-D$'$-NS$'$ configuration, which we claimed corresponded to $W \sim
Q_1 \tilde{Q}^1 Q_2 \tilde{Q}^2$. Performing the mirror transformation
of $SL(2,\IZ)$ followed by rotations and exchanging orders of branes,
as in \hw, we come back to exactly the same configuration, so we claim
this theory is self-mirror (like the corresponding $N=4$
theory). However, this may also be seen directly from the known $N=2$
mirror for the theory with no superpotential \refs{\bhoy,\ahiss}. This
mirror is a $U(1)$ theory with two flavors $q_i,\tilde{q}^i$, 
two singlets $S_i$, and a
superpotential $W = S_1 q_1 \tilde{q}^1 + S_2 q_2 \tilde{q}^2$. Since
the $S_i$ are identified with $Q_i \tilde{Q}^i$ \ahiss, the additional
superpotential in the mirror theory is $W \sim S_1 S_2$, and then
integrating out $S_1$ and $S_2$ we indeed get the same quartic
superpotential as in the original theory.

\subsec{Branes and Superpotentials in 4D $N=1$ Supersymmetric Gauge Theories}

Much of the discussion of this section can be generalized also to
brane constructions of 4D $N=1$ field theories, which are different
from the constructions described above only in the fact that all the
D-branes stretch along the $x_3$ direction as well. One obvious
difference is that the vector multiplet no longer contains any
scalars, so there are no longer any Coulomb branches, but the analysis
of the Higgs branches described above is essentially the same. In
particular, also in brane constructions of 4D $N=1$ theories with
D6-branes in different orientations we expect to get quartic
superpotentials, as described above. In the 4D theory, the coefficient
of this superpotential carries (in the UV) a negative mass dimension,
and we expect it to be of the order of one over the string
scale. However, the presence of the superpotential still lifts some of
the Higgs branches, and the field theory with the superpotential still
describes correctly the brane construction.

There are two types of instanton effects in the brane constructions of
the 4D $N=1$ theory, related to the two types of instanton effects in
3D $N=2$ theories described above. First, there are the standard gauge
theory instantons, which (in the zero size limit) look like D0-branes
inside the D4-branes, and which are expected to generate
superpotentials lifting the Higgs branches in some cases (as in
\ads). Then, there are the ``Higgs branch instantons'' similar to the
ones which appeared in 3D. When two D4-branes stretch between D and
D$'$ 6-branes, with no other branes nearby, the effect of an Euclidean
string stretched between them is expected to lift this branch, in a
way which we cannot describe in the original field theory, since it
depends on parameters which are not visible there
(in the 4D case this instanton 
is not related by any duality to the standard
instantons). As in the 3D case discussed above, the lifting of these
branches is consistent with the fact that they do not appear in the
corresponding field theories. The full analysis of the brane
configurations is, however, complicated by the fact that the standard
4D instanton effects are not easily visible in these configurations.

As we did for mirror symmetry in the previous section, we can also
generalize the constructions of Seiberg dualities \sem\ from branes
\egk\ (see also \refs{\oogvaf,\ejs,\bh,\telaviv,\ahnoh})
to the configurations with D-branes at different
angles. However, again, all we get is a deformation of the known
dualities by an operator quartic in the quarks of the original theory,
which we can identify also in the dual theory. For instance, beginning
with the NS-D$^n$-(D$'$)$^m$-NS$'$ configuration, which we claimed was
described by $W \sim (Q_i \tilde{q}^j) (q_j \tilde{Q}^i)$ (where $i$
goes over the D-quarks and $j$ goes over the D$'$-quarks), we can move
all D branes to the left and all D$'$ branes to the right without
changing the low-energy theory, and then exchange
the NS and NS$'$ 5-branes as in \refs{\hw,\egk}. In the brane
construction, after the exchange we
find an $SU(N_f-N_c)$ theory 
with $n+m$ flavors, with $n^2$ mesons coupling to $n$ of the dual
quarks and $m^2$ mesons coupling to the other $m$ dual quarks. This is
exactly what we find also just by adding the operator corresponding to
the superpotential $W$ above to the standard Seiberg dual \sem, since
it just gives a mass to the ``off-diagonal'' mesons. Integrating out
these mesons we find a quartic superpotential also for the dual
quarks, which we expect to arise in the dual brane configuration in
the same way discussed above, and which lifts the dual off-diagonal
Higgs branches (that do not exist in the original theory). The same
discussion applies also to the 3D $N=2$ configurations using the
duality described in \methree\ (see also \karch). The microscopic
definition of this duality is not yet known, but it correctly matches
the effective IR theories describing the ``electric'' and ``magnetic''
theories.

\subsec{Chiral symmetry}
\subseclab{\chiral}

It was conjectured in \bh, in the context of 4D $N=1$ gauge theories
realized through type IIA superstrings, that non-Abelian chiral
symmetry is manifested in a particular configuration of the
branes\foot{The axial $U(1)$ is usually directly visible in the brane
constructions -- when all D-branes have the same orientation it is
just the difference between $SO(2)_{45}$ and $SO(2)_{89}$.}. This
happens when a D6-brane parallel to a NS 5-brane coincide. The gauge
symmetry on the worldvolume mutual to both types of branes, when
there are $n$ coincident D6-branes, seems to be enhanced from
$SU(n)_V$ to $SU(n)_L\times SU(n)_R$ when the position of the
D6-branes coincides with the position of the NS brane. The same
construction obviously holds also for the 3D $N=2$ theories discussed
in this paper, when D and NS$'$ 5-branes (or D$'$ and NS 5-branes)
come together. In this case the D5-branes can actually split along the
NS 5-brane, as discussed in the next section, and the implications of
this for chiral symmetry are discussed at the end of section 4.

Naively, we would not expect to find the non-Abelian chiral symmetry
in the brane constructions, even though it exists in the low-energy
field theory. This is because, as described above in detail, the
massive $\phi$ fields couple to the quarks in a way which breaks the
chiral symmetry. In particular, the quartic superpotentials we
described above, which we get when integrating out these $\phi$
fields, are obviously not invariant under the chiral
symmetry. However, in the particular configurations where all the
D$'$ branes are to the left of all the D branes, we found that no such
quartic superpotential exists. These are exactly the configurations
where we can move (in $x_6$) all the D$'$ branes to intersect with the
NS branes, and all the D branes to intersect the NS$'$ branes, without
encountering any phase transitions, and thus, this is the only case
where the arguments of \bh\ for chiral symmetry hold. Therefore, the
superpotentials we find are consistent with the conjecture of
\bh. However, it is still not clear why the
quartic superpotentials breaking the chiral symmetry are not generated
in these configurations, or, equivalently, why the apparent coupling
of the $\phi$ fields to the quarks no longer breaks the chiral
symmetry. The consistency of the conjecture of \bh\ seems to require
that either the $\phi Q \tilde{Q}$ coupling goes away when the
``flavor'' D-branes intersect the NS 5-branes, or that the mass of
$\phi$ goes to infinity in this limit. Both assumptions are consistent
with our results here, but it is not clear why they should be true. This
issue is obviously related to the phase transition when a D brane
passes a NS$'$ brane in $x_6$, since the chiral symmetry is claimed to
be manifest at the point of this phase transition. This phase
transition is discussed in the next section.

\subsec{Other effects of ``hidden parameters''}

In \S2.4 we saw that the relative $x_6$ positions of the D5-branes,
which were ``hidden parameters'' in the $N=4$ theory discussed in \hw,
actually have an effect on the low-energy field theory in some
cases. In the $N=2$ configurations we discuss here, there is also
another ``hidden parameter'' that could affect the low-energy theory,
which is the relative $x_6$ position of D and NS$'$ branes (or of D$'$
and NS branes). Recall that in \hw\ it was discovered that moving a D
brane through a NS brane led to a phase transition in which a new
D3-brane was generated, enabling the low-energy field theory to stay
the same after the transition. Thus, the relative $x_6$ positions of D
and NS branes (or of D$'$ and NS$'$ branes) do not affect the
low-energy field theory.

However, for this phase transition to occur it was crucial that the D
and NS branes had to intersect when they were interchanged, and this
is no longer true for D and NS$'$ branes (though it is true for any
other angle, for which the branes are not parallel in the 45-89 complex
plane). D and NS$'$ branes can be interchanged by moving them around
each other in the $x_4$ or $x_5$ directions, and then it seems clear
that no phase transition can occur when they are interchanged, since
they do not have to pass near each other. There is also no linking
number argument for a phase transition in this case. Thus, the
apparent conclusion would be that there is no phase transition when D
and NS$'$ branes are interchanged, and then their relative $x_6$
position obviously affects the low-energy dynamics (since we have a
massless quark when the D is between the NS and the NS$'$ but not
otherwise).

However, this conclusion might be too naive. When we move the branes
around each other as described above, we are giving the quarks arising
from the D5-branes a mass. If we want the D5-brane and the NS 5-brane
to really go around each other smoothly, the distance between them
should be larger than the string scale, but in this case the quark
mass is also very large, and it can no longer be included in the
low-energy effective theory (since there are many other states whose
mass is of the order of the string scale in the brane
construction). So, we cannot really follow what happens to the quarks
in the low-energy field theory by taking the branes around each other,
and a phase transition might still occur if the branes are
interchanged at a distance smaller than the string
scale. Unfortunately, it does not seem possible to analyze this
transition by string theory methods, nor by the methods used in
\oogvaf\ to study the phase transition of \hw. The issue of this phase
transition, and in general of the $x_6$ dependence of the low-energy
field theory as described above, deserves further investigation.

\fig{A phase transition through ``hidden parameters''. There are $N$ 
D3-branes stretched between NS and NS$'$ 5-branes, and $N'$ D3-branes 
stretched
between NS$'$ and NS 5-branes. A D 5-brane (denoted by ``X'') can move
in the $x_6$ direction to give different massless matter
contents.}{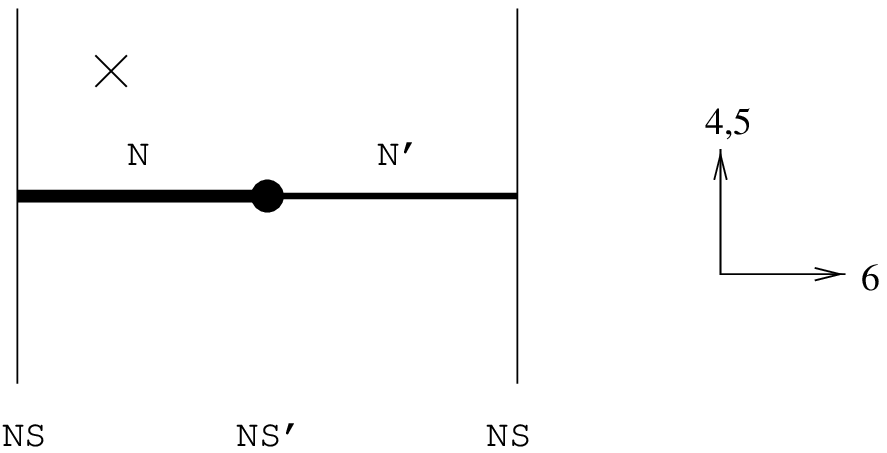}{8 truecm}
\figlabel\parad

As a specific example let us consider a configuration of three
NS 5-branes as depicted in figure \parad. The two external 5-branes are NS
and the 5-brane in the middle is NS$'$.
There are $N$ D3-branes between the left NS brane and the NS$'$ brane, and
$N'$ D3-branes between the right NS brane and the NS$'$ brane.
There is also a D 5-brane located at arbitrary positions in $x_{3,4,5,6}$.
Denote the $x_6$ position of the D 5-brane by $z$ and the $x_6$ position of
the NS$'$ brane by $t$.
The 3D gauge group is $U(N)\times U(N')$.
The D brane gives rise to a (generally massive) quark (and perhaps
other states as well).
Moving the D 5-brane in the 3 and 45 directions corresponds to changing the
bare real and complex masses, respectively, of the quark.
In particular, when the D brane touches the D3-branes, there is a massless
quark. The matter content depends, however, on the ordering of $z$ and $t$.
For $z<t$ we have a massless quark in the $(N,1)$ representation of
the gauge group while for $z>t$ we have a massless quark in the
$(1,N')$ representation of the gauge group.  There is a smooth process
which interpolates between the two regions.  Starting with a massless
quark in the first region, we can move the D5-brane in the 45
directions, change the $x_6$ ordering and move back to get a massless
quark in the second region. There is no D3-brane created in the
process as the NS$'$ and D branes can go around each other, but the
massless matter content is obviously changed. From a field theory
point of view such a process, of a change in the representation of a
field, would be very surprising, but, as discussed above, it seems
that we cannot really discuss such processes in the low-energy field
theory. Presumably, this type of process indicates that there are
massive states in the theory whose mass depends on the ``hidden''
$x_6$ positions.

\newsec{``Polymeric'' 5-branes and Five Dimensional Field Theories}

In this section we will construct five dimensional gauge theories, and
other five dimensional superconformal field theories,
using D5-branes and NS 5-branes in type IIB string theory. 
The construction we use will be a generalization of the construction
of \hw\ in 3 dimensions, and of \newwitten\ in 4 dimensions, but we
will use slightly different conventions for reasons that will become
clear when we relate our results here to three dimensional gauge
theories in section 4.

\fig{A D5-brane which ends on a NS 5-brane. The left side
describes the naive configuration, and the right side the correct
configuration, which implements conservation of charge at the vertex.}
{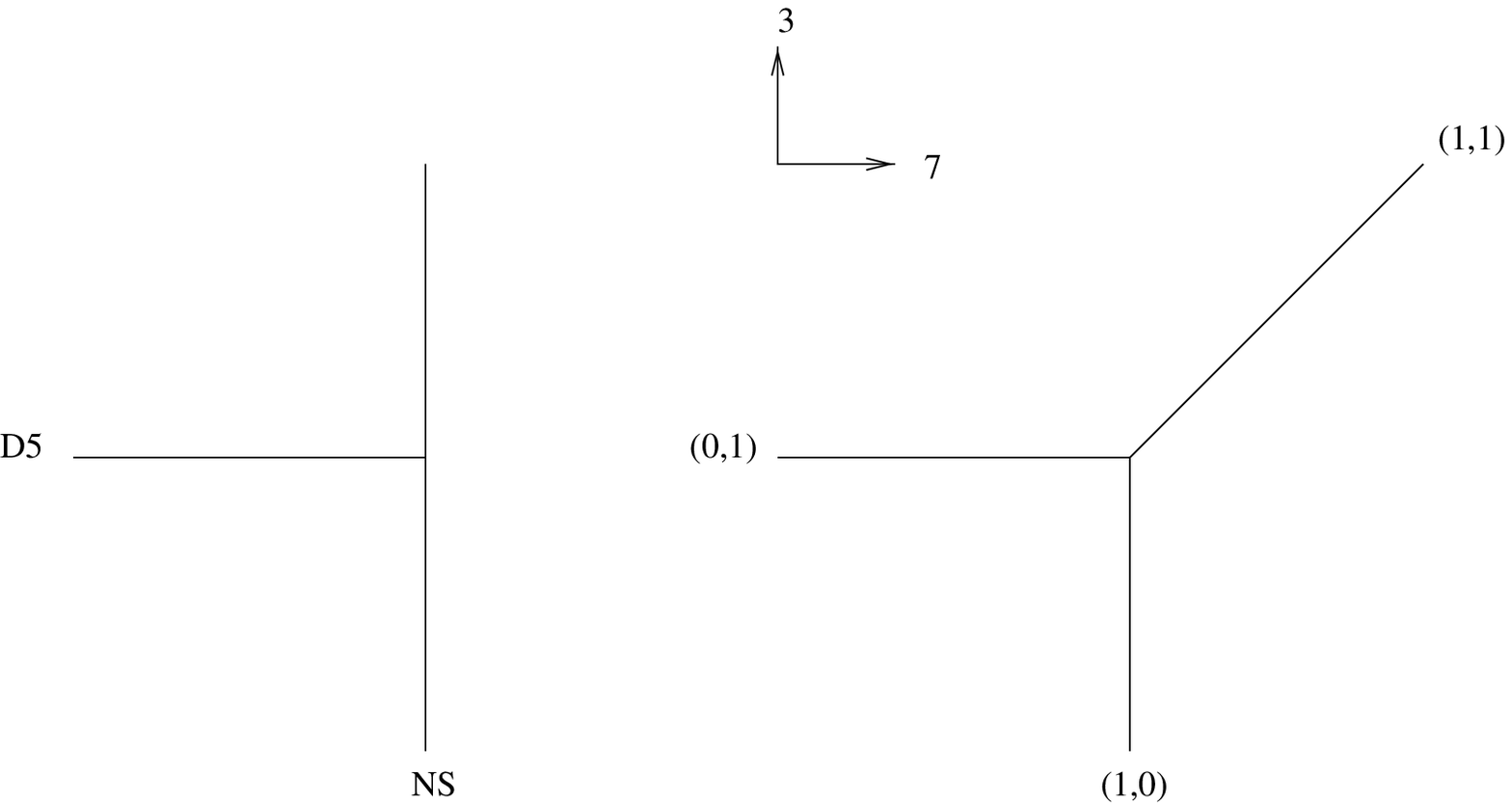}{15 truecm}
\figlabel\NSD

Consider a NS 5-brane along the 012389 coordinates and a D5-brane along
the 012789 coordinates which ends on a it, as drawn on the left
side of figure \NSD\ (this is a naive description of the situation
which will be improved below).
This system is T-dual to any D-brane ending on a NS 5-brane, like in
the configurations studied in \hw.
Thus, it is easy to check that the
supersymmetry is broken to 1/4 of the original supersymmetries. This
means that we are dealing with minimal ($N=1$) supersymmetric six 
dimensional gauge theories, or five dimensional gauge theories if we
make some of the 5-branes finite (as we did for the D3-branes
above). These theories have an $SU(2)_R$ global symmetry, which is a
double cover of the $SO(3)$ rotation group in the $x_4,x_5$ and $x_6$
directions. The boundary of the D5-brane is a 4-brane which
propagates in $5+1$ dimensions, and thus has the behavior of a particle
in $1+1$ dimensions. There are four scalars which parametrize the
position of the NS 5-brane in 4567 space, but it is clear from
symmetry arguments that
only the $x_7$ coordinate 
is affected by the presence of the D5-brane, and it will
depend on the position of the D5-brane in the $x_3$
direction (the analysis here is a generalization of the analysis of
\newwitten\ for D4-branes ending on NS 5-branes). 
For large $x_3$, the classical equation
which governs this dependence 
is the Laplace equation in one spatial dimension :
\eqn\laplace{\nabla^2x_7=\delta(x_3),}
where $x_3=0$ is the position of the D5-brane inside the NS brane.
The general solution of this equation is 
\eqn\linear{x_7= {1\over 2}|x_3| + c x_3 + d,}
where $c$ and $d$ are constants. We will choose the constants so that
for large negative values of $x_3$, we have a standard NS(012389)
5-brane at $x_7=0$ -- this determines $c={1\over 2}$ and $d=0$. 
At this point it seems like the supersymmetry is broken, 
since we started with
a NS 5-brane along the $x_3$ coordinate, and we end up (for $x_3 > 0$)
with a brane
which is diagonal in the $x_3-x_7$ plane. How can this be possible?
Recall that in fact, in the type IIB string theory, there are bound
states of $p$ NS 5-branes and $q$ D5-branes (for $p$ and $q$
relatively prime), called $(p,q)$ 5-branes, which behave just like
standard 5-branes (and, in fact, are related to them by $SL(2,\IZ)$
U-duality transformations). Charge conservation does not really allow
a D5-brane to end on a NS 5-brane -- instead, at the intersection
point the two 5-branes merge together to form a $(1,1)$ 5-brane, as
drawn on the right side of figure \NSD. Charge conservation does not
constrain the configuration any further. However, if we want the
``new'' brane coming out of the intersection point not to break the
supersymmetry any further, it must be oriented at a 45 degree angle in
the $x_3-x_7$ plane, exactly as \linear\ implies.
In the same
way general vertices of $(p,q)$ 5-branes may be drawn,
subject to charge conservation
(a similar
discussion for $(p,q)$ strings appears in \asy), 
where the angle $\theta$ of each $(p,q)$ 5-brane in
the $x_3-x_7$ plane satisfies $\tan(\theta)=p/q$ so as to preserve the
remaining supersymmetry.

Naively, we may expect corrections to equation \linear, but there
are no objects which can contribute to the metric of the
configuration. Any corrections would also break the remaining
supersymmetry. This is consistent with the expectations from field
theory (when we use these configurations to construct 5D field
theories, as described below), since in five dimensions there are no
instanton corrections to the metric.

We can easily extend this analysis to cases in which there is more
than one D5-brane ending on a (asymptotically) NS 5-brane from both
sides.  Let $a_i$, $i=1,\ldots,m$ be the $x_3$ positions of the 
D5-branes ending from the left on the NS 5-brane, and $b_j$,
$j=1,\ldots,n$ be the $x_3$ positions of D5-branes ending from the
right on the NS 5-brane. The solution of the generalization of
\laplace\ then takes the form
\eqn\multilinear{x_7={1\over 2}(\sum_{i=1}^m|x_3-a_i|-
\sum_{j=1}^n|x_3-b_j|)+c x_3 + d.}
For $c={{m-n}\over2}$ and $d=0$ this equation has the interpretation of a
5-brane which starts far away as a
NS 5-brane, and changes its charge, and its $x_3-x_7$ angle, 
in places where a D5-brane ends on it.
The change of charge is dictated by a conservation 
law which states that the sum of charges be zero at each vertex point,
while the angle is determined by supersymmetry (or by \multilinear).
A condition for a NS brane to stay NS for large $x_3$ is that $m=n$.
In all other cases the 5-brane looks different at the far ends.
The piecewise linear function \multilinear\ is reminiscent of the five
dimensional gauge coupling functions described in
\refs{\seibergfive,\morsei,\dkv,\fived}, 
and it is in fact closely related to them,
as described in the next section.

\subsec{Pure $SU(2)$ gauge theory}

Next, we turn to the construction of five dimensional gauge
theories using the branes. As in \refs{\hw,\newwitten}, a
configuration of $N_c$ parallel D5-branes of finite extent, stretched
between NS 5-branes, is expected to give a $U(N_c)$ (or $SU(N_c)$) 5D
$N=1$ supersymmetric gauge theory.  
For a given configuration of branes there can
be two types of deformations : one type which does not change the
asymptotic form of the 5-branes, and a second type which changes the
asymptotic form of the 5-branes.  As in \newwitten, we interpret the
first type of deformation as a change in the dynamical moduli of the
system, while the second type is interpreted as a change in the
parameters which define the field theory.

\fig{Pure $SU(2)$ gauge theory in five dimensions. Horizontal lines represent
D5-branes, vertical lines represent NS 5-branes, and diagonal lines
at an angle $\theta$ such that $\tan(\theta)=p/q$ represent $(p,q)$
5-branes. Figure (a) shows a generic point on the Coulomb branch,
figure (b) shows a point near the origin of moduli space, and figure (c)
corresponds to the strong coupling fixed point.} {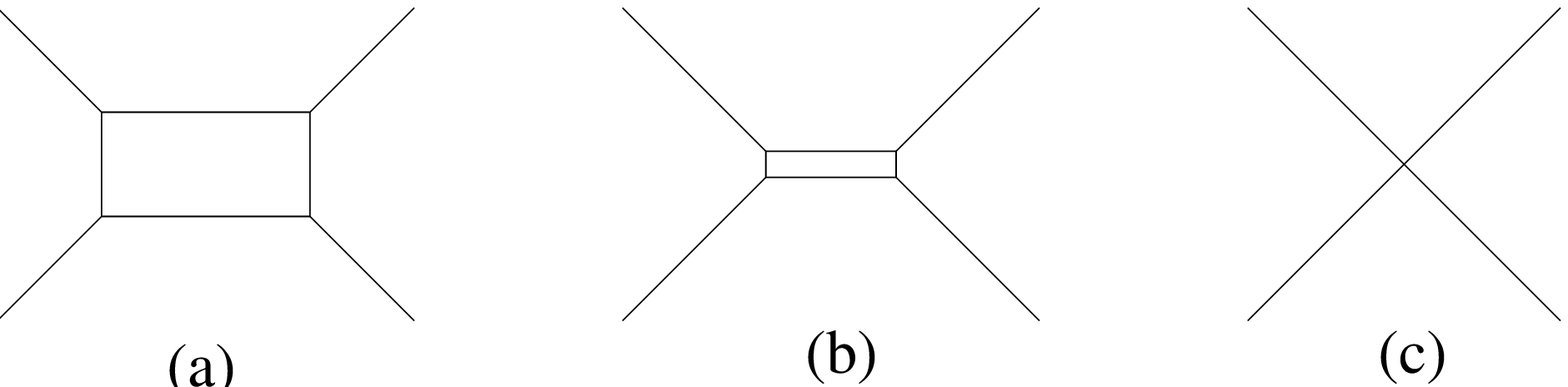}{13 truecm}
\figlabel\SUtwozero

Let us begin with the simplest configuration of two
parallel D5-branes. According to the previous discussion, 
this configuration
actually looks like the left hand side of figure \SUtwozero(a), where
we chose (arbitrarily) specific orientations for the outgoing
branes. Apriori this seems to represent a $U(2)$ gauge theory, and one
might expect also additional contributions from the finite segments of
the NS 5-branes, which also correspond to 5D theories at low
energies. However, as in \newwitten, it is easy to see that there is
only one deformation of this configuration which does not change the
asymptotic forms of the branes in figure \SUtwozero, corresponding to
going from figure \SUtwozero(a) to \SUtwozero(b). Thus, the
configuration \SUtwozero(a) has only one real massless scalar, which
(using the 5D supersymmetry) necessarily corresponds to having just one
massless vector multiplet (and no massless hypermultiplets). We will
denote this scalar by $\phi$; it can be chosen to correspond to the
distance between the two D5-branes. We expect to find an unbroken
$SU(2)$ gauge symmetry when the two D5-branes come together, as in
\SUtwozero(b). This is consistent with the
obvious $\phi \to -\phi$ symmetry, which implies that the vector
multiplet we see is indeed in the Cartan subalgebra of an $SU(2)$
gauge theory. Thus, we suggest that the configuration \SUtwozero(a)
corresponds to a generic point on the Coulomb branch of the 5D $N=1$ pure
$SU(2)$ gauge theory.

We can check this conjecture by examining various properties of this
configuration. As in \refs{\hw,\newwitten}, it is natural to identify
the bare gauge coupling $1/g_0^2$ with the length of the two D5-branes
at the point in moduli space where they come together (up to a
constant involving the string coupling). The Coulomb branch of the
$SU(2)$ gauge theory has a prepotential of the (exact) form \seibergfive\
\eqn\prepot{{\cal{F}} = {1\over {2g_0^2}} \Phi^2 + {c\over 6} \Phi^3,}
where $c = 16 - 2N_f$ arises at one loop. The masses of electrically
charged BPS saturated states are proportional to $\phi$, while those
of magnetically charged BPS saturated states (which are strings in
5D) are proportional to ${\del {\cal{F}} \over {\del \phi}}$. In
the configuration of figure \SUtwozero, the electrically charged
particles (W bosons) arise from strings between the two D5-branes, so
their mass is naturally proportional to $\phi$. 
As in \hw, the magnetic monopole
may be identified with a D3-brane spanning the rectangle between the D
and NS 5-branes (thus giving rise to 5D strings), since a D3-brane
looks like a magnetic monopole from the point of view of any 5-brane
it ends on. The tension of these strings is proportional to the area
of the rectangle, which is (up to constants) $\phi/g_0^2 + \phi^2$,
in agreement with \prepot\ (up to multiplicative redefinitions of
$g_0$ and $\phi$).

The $SU(2)$ gauge theory with no flavors is argued to have a
non-trivial fixed point as $g_0 \to \infty$ \seibergfive.  In the brane
construction, this corresponds to figure \SUtwozero(c) (choosing
$\phi=0$). Thus, we claim that this fixed point is the same as the one
corresponding to the intersection of $(1,1)$ and $(1,-1)$ 5-branes at
90 degrees (with five dimensions common to both branes). Note that
this configuration is not equivalent in any way to a configuration of
a NS 5-brane and a D5-brane intersecting at 90 degrees, which has no
Coulomb branch. 

The brane construction suggests several deformations
of this fixed point
(here we discuss changes of parameters of the 5D field theory,
which correspond to the asymptotic locations/orientations of the
5-branes, as opposed to local changes which correspond to 5D fields
such as $\phi$).

The obvious deformation
is increasing $1/g_0^2$, going from figure \SUtwozero(c) to
\SUtwozero(b). This deformation is visible from the field theory point of view
\seibergfive\ and corresponds to turning on a finite $1/g_0^2$ in a
$SU(2)$ gauge theory, which then flows to a trivial IR fixed point.

A deformation in the opposite direction, which might be
called a continuation of the field theory to negative values of
$1/g_0^2$, leads to a configuration which looks like a 90 degree
rotation of figure \SUtwozero(b). This looks like an $SU(2)$ gauge
theory with coupling $|1/g_0^2|$, but now the $SU(2)$ gauge symmetry
comes (at least naively) from the NS 5-branes instead of from the
D5-branes.

There is one other deformation we can do at the infinite
coupling fixed point without breaking the supersymmetry, which corresponds to
moving the two branes which intersect there apart in the $x_{4,5,6}$
directions. After this deformation there are no longer any massless
five dimensional fields. When the two 5-branes approach each other, the
lowest lying 5D states naively correspond to membranes whose tension is
proportional to the distance between the two 5-branes. These arise from
D3-branes ending on both branes (these are the only branes which can
do so). Thus, from this point of view the $SU(2)$ strong coupling
fixed point has ``tensionless membranes''. In the
construction of these theories via M theory on Calabi-Yau manifolds
\refs{\morsei,\dkv,\gms}, such deformations correspond to blowing 
up 3-cycles (and the membranes arise from 5-branes wrapped around
these 3-cycles)\foot{The number of such 3-cycles in some particular
cases was computed in \cggk.}. 
It may be possible to translate our constructions
directly into those of \refs{\morsei,\dkv,\gms} by translating
5-branes into geometric singularities as in \oogvaf.

In our construction of the $SU(2)$ theory we made an arbitrary choice
of the orientation of the outgoing 5-branes. Instead of choosing
$(1,1)$ and $(-1,1)$ 5-branes for large values of $x_3$, we could have
chosen general $(p_1,q_1)$ and $(p_2,q_2)$ 5-branes, and most of the
discussion above would have remained unchanged (at least as long as
all the branes appearing have $(p,q)$ relatively prime). The Coulomb
branches corresponding to all these constructions are obviously
isomorphic, but it is not clear to us if the strong coupling fixed
points corresponding to these different constructions are the same or
not (i.e. whether they have the same operators and OPEs). The theories
might differ by irrelevant operators. Recall that,
as discussed in \seibergfive\ and reviewed in \S3.4 below, 
the $SU(2)$ pure gauge theory
is known to have two different possible strong coupling limits, called
$E_1$ and $\tilde{E}_1$
(corresponding to different discrete ``theta parameters'' \dkv). One
property of the strong coupling fixed points which we can try to read
off from the brane configuration is their global symmetries, though it
is not clear if all the non-Abelian factors of the global symmetries
must be manifest in this type of construction. As discussed in
\S3.2 below, 
the rank of the global symmetry for any choice of asymptotic
branes is one, but generically there is no sign of the enhanced
non-Abelian gauge symmetry that corresponds to the $E_1(=SU(2))$ fixed
point \seibergfive.

\fig{Another construction of the pure $SU(2)$ gauge theory. Figure (a)
shows a point near the origin of the Coulomb branch, figure (b) shows
the strong coupling fixed point, and figure (c) shows the Coulomb
branch emanating from this strong coupling fixed point.}
{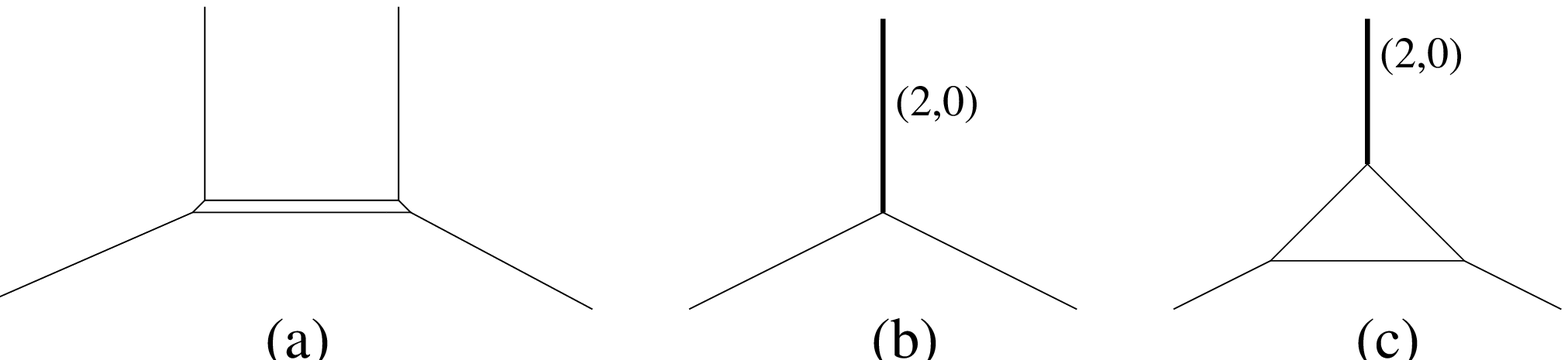}{15 truecm}
\figlabel\Eonetheory

In one particular configuration, described in figure \Eonetheory(a),
where we choose both of the asymptotic branes for large $x_3$ to be NS
5-branes, we do see an enhanced $SU(2)$ global symmetry at the strong
coupling fixed point, described by figure \Eonetheory(b), 
where the two NS 5-branes overlap.
It is not clear if this $SU(2)$ global symmetry
exists (but is hidden) for other constructions of the strong coupling
fixed point, or if the other constructions describe different fixed
points. In the latter case our constructions give many new 5D $N=1$
superconformal field theories, all of which have a one dimensional
Coulomb branch similar to that of the $SU(2)$ pure gauge theory.
Another difference between the brane constructions with different
asymptotic branes is that only in the original construction of figure
\SUtwozero\ do we see the deformation corresponding to having
membranes with small tensions. 
Again, this might indicate a difference between the
different fixed points, or that this deformation is just not visible
in the other brane constructions of the same fixed point. We have not
been able to determine whether the various fixed points are the same
or not.

\subsec{General properties of 5D fixed points from branes}
\subseclab{\general}

As discussed above, it seems possible to get five dimensional SCFTs
from the low-energy theory at the
intersection of several half 5-branes (each of which has some
charges $(p,q)$). In this section we discuss the general properties of
such fixed points which appear in the brane constructions.

The obvious generalization of the fixed points described in the
previous section is to a configuration of $n$ half 5-branes which all
emanate from the same point. We claim that each such configuration
corresponds to a five dimensional SCFT (though in some cases, like
$n=2$, this theory will be trivial). We will denote the charges of the
half 5-branes, when they are all oriented towards the intersection
point (i.e. their orientation 6-form is $dx_0 \wedge dx_1 \wedge dx_2
\wedge dx_8 \wedge dx_9$ times a vector in the direction of the
intersection point), by $(p_i,q_i)$ ($i=1,\cdots,n$). Charge
conservation obviously requires $\sum_{i=1}^n p_i = \sum_{i=1}^n q_i =
0$, and the angles of the branes are related to their charges by the
condition of preserving the remaining supersymmetry. Note that these conditions
are not $SL(2,\IZ)$ invariant (the S generator of $SL(2,\IZ)$ acts on
them in the same way as a 90 degree rotation in the $x_3-x_7$ plane,
but there is nothing similar for the T generator), so we cannot do an
$SL(2,\IZ)$ transformation taking, say, $(p_1,q_1)$ to $(0,1)$ without
(generally) changing the fixed point theory. We will assume that all
the $(p_i,q_i)$ are relatively prime. A general 5-brane for which
$(p,q)=m(p_0,q_0)$
may be viewed as a collection of $m$ $(p_0,q_0)$ 5-branes (and will be
drawn as a wide line in our figures).

This construction defines (implicitly)
the fixed point theory corresponding to particular values of
$(p_i,q_i)$. There are two obvious types of questions we can ask about
these fixed points in the brane configurations. First, we can ask what
are the Coulomb branches coming out of these fixed points,
generalizing the $SU(2)$ Coulomb branch described in the previous
section. These branches correspond to deformations of the 5-branes
near the intersection point which do not change the asymptotic
behavior of the branes. They may include several parallel D5-branes,
as in the previous section, in which case we can interpret them (at
least naively) as corresponding to a non-Abelian gauge theory arising
from these D5-branes. This will be discussed in the next
section. There could also be Coulomb branches which have several
parallel $(p,q)$ branes, which we could also interpret in the same
way, or branches which have no parallel branes at all. An example of
the latter possibility is the strong coupling limit of the
$SU(2)$ $N_f=0$ theory (described in the previous
section) with two asymptotic NS 5-branes, whose Coulomb branch is
described in
\Eonetheory(c). In such cases we cannot interpret the W bosons as
corresponding to particular strings between 5-branes, and they
correspond to more general states (which, in principle, are
combinations of states from all the 5-branes which have one finite
direction). A general analysis of the Coulomb branch emanating from a
particular fixed point (and, in particular, a computation of its
dimension as a function of the charges $(p_i,q_i)$) 
should be possible, but we will not perform it here, and
discuss only some specific cases.

The other properties of the fixed points which we can read off from the
brane constructions are their deformations and global symmetries (which
are related since some of the deformations may be thought of as
turning on background global vector fields). There seem to be two
types of deformations which do not break the supersymmetry. First, we
can move the branes in the $x_3-x_7$ plane without changing their
orientation. The line corresponding to each $(p_i,q_i)$ 5-brane is
given by an equation of the form $p_i x_7 + q_i x_3 = c_i$ for some
real number $c_i$, so naively we have $n$ real parameters
corresponding to the positions of the branes. However, if the branes
are not all parallel (in which case there is no interesting 5D
fixed point), two of these parameters correspond to setting the origin
of the $x_3$ and $x_7$ coordinates, and another is set by the
requirement that $\sum_{i=1}^n c_i = 0$. This requirement is obvious
if all 5-branes are to intersect at one point, but it holds also in
more general configurations which involve also finite 5-brane
segments, since it holds at each intersection point separately. Thus,
there are only $n-3$ real parameters corresponding to deformations of
the fixed point theories. These parameters are the scalar components
of vector multiplets inside the 5-branes. Naively, there is (at least)
a $U(1)^n$ gauge symmetry on the 5-branes, which may be interpreted as
(part of) the global symmetry of the fixed points, but the discussion
above shows that only a $U(1)^{n-3}$ subgroup of this acts
non-trivially on the fixed points, and the deformations described
above can be viewed as background vector fields for these global
symmetries. When $m$ of the 5-branes have the same $(p_i,q_i)$ and
$c_i$, the global symmetry is enhanced to (at least) $SU(m)$, but its
rank always remains $n-3$ (and the deformations correspond in general
to the Cartan subalgebra of the global symmetry group).

The second type of deformation which is visible in the brane
constructions corresponds to moving branes away in the $x_4,x_5$ or
$x_6$ directions. Such a deformation appears whenever two of the
branes have opposite $(p_i,q_i)$ and $c_i$, and then we can join them
together to one infinite (as opposed to semi-infinite) 5-brane and
move them off. This type of deformation is charged under the $SU(2)_R$
global symmetry (which we identified with $SO(3)_{456}$), so it cannot
be described by a background vector multiplet. Instead, its form is
like that of a \FI\ term, but its interpretation as such is not clear
since the $U(1)$ factors of the 5D gauge group seem to be projected
out, as described above. Note that deformations of this sort are
generally possible only at the strong coupling fixed points, and not
at generic points on the Coulomb branch, so it is not clear that they
should have a field theory interpretation. When approaching the fixed point
from the direction of a deformation of this sort, the 5D theory seems to
have membranes whose tension goes to zero at the fixed point.

\subsec{Generalizations to $SU(N_c)$ gauge theories with $N_f$ flavors}

The generalization of our construction of the $SU(2)$ gauge theories
to any value of $N_c$ is straightforward. Instead of two parallel
D5-branes, we have $N_c$ parallel D5-branes, still between two other
branes, which we can choose to have charges $(p_1,q_1)$ and
$(p_2,q_2)$ for large values of $x_3$, and then for large negative
values of $x_3$ they will have charges $(p_1,q_1-N_c)$ and
$(p_2,q_2+N_c)$. The $S_{N_c}$ Weyl symmetry corresponds, as usual, to
permuting the D5-branes, and the $U(1)$ part of the naive gauge group
is projected out as in the $SU(2)$ case (and as in \newwitten).

\fig{A point on the Coulomb branch of the pure $SU(3)$ gauge theory.}
{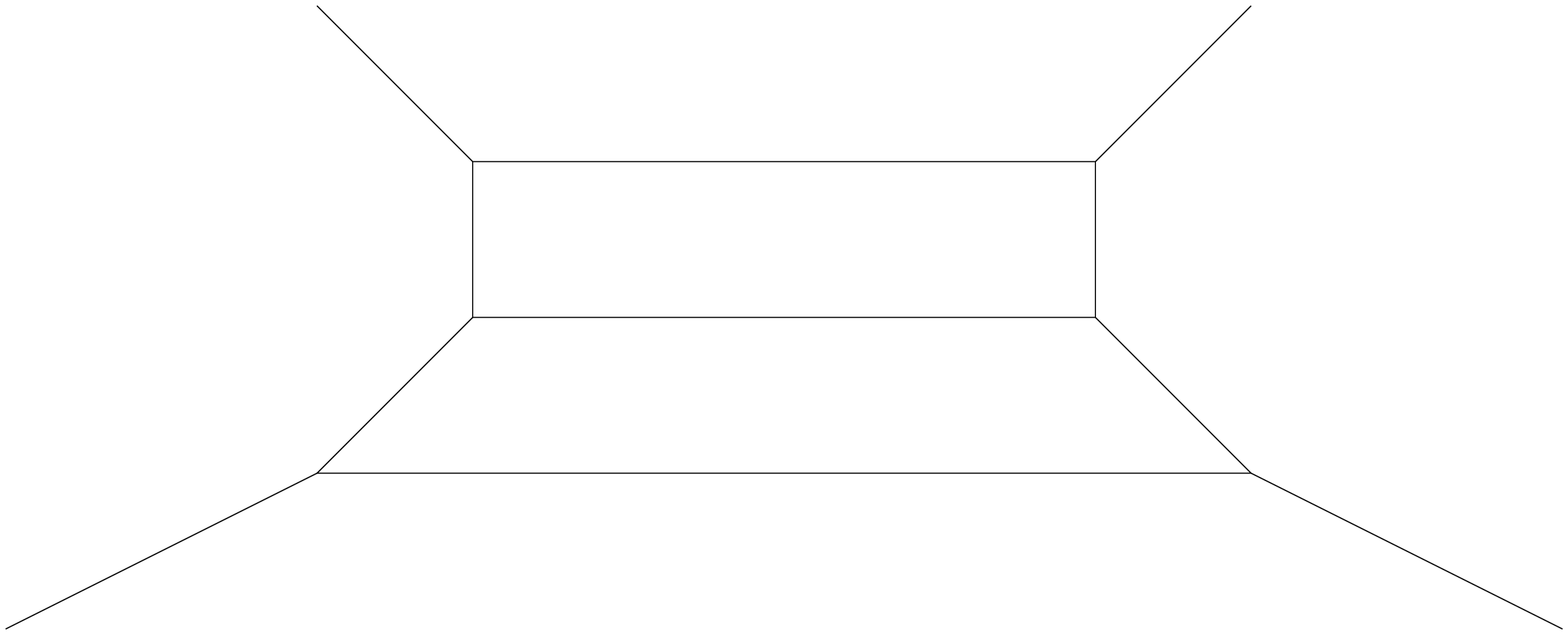}{10 truecm}
\figlabel\SUthree

For example, a particular construction of the pure $SU(3)$ gauge
theory is depicted in figure \SUthree. The prepotential for this
theory was computed in \fived. As for $SU(2)$, we can identify
$1/g_0^2$ with the length of the D5-branes when they are all
together. The 5D monopoles, whose tensions are given by linear
combinations of derivatives of the prepotential, again correspond to
D3-branes stretching over finite surfaces in the $x_3-x_7$ plane. The
area of these surfaces agrees with the known formulas for the monopole
tensions (it is clear from the figures and from the generalizations of
\multilinear\ that they are always piecewise quadratic in the $x_3$
positions of the D5-branes, which are related to the Cartan scalars
$\phi_i$). However, in these configurations we can no longer simply
identify the W boson masses with the distances between the
D5-branes. This is not too surprising since the other finite branes in
these constructions should be just as important as the D5-branes. For
instance, at the $SU(2)$ strong coupling fixed point of figure
\SUtwozero(c), there is a symmetry between the D and NS branes. The
``polymers'' describing the gauge theories which
we find here are analogs of
the Seiberg-Witten curve \swfour\ of 4D $N=2$ gauge theories, which
was constructed in a similar way in
\newwitten.

As we did in the $SU(2)$ case, we can take the strong coupling limit
of all these constructions, and obtain strong coupling fixed points
for the pure $SU(N_c)$ gauge theories, of the general form discussed
in \S3.2. All these fixed points (corresponding to different
asymptotic branes) are described by an intersection of four half
5-branes, so their global symmetry is of rank one. Generally, only a
$U(1)$ global symmetry is visible, except when we choose two of the
outgoing branes to be parallel, and then the visible global symmetry
is $SU(2)$. Also, as in the $SU(2)$ case, there is at least one
particular choice of asymptotic branes for which a \FI-like
deformation is visible at the strong coupling fixed point
(corresponding to moving the branes apart in
$x_4, x_5$ or $x_6$), given by
$(p_i,q_i)=\{(-1,N_c-1),(1,1),(1,1-N_c),(-1,-1)\}$.

Next, we can add quarks (charged hypermultiplets) to these
theories. As in previous brane constructions, there are two ways to do
this -- either by adding D7-branes in the 01245689 directions, which
intersect the D5-branes giving rise to the gauge group (as in \hw, but
D7-branes affect the asymptotic space-time geometry so their analysis is
more complicated), or by adding semi-infinite D5-branes parallel to
the ones we have (as in \refs{\hw,\newwitten}). We will choose the second
construction since its analysis is simpler (though the Higgs branches
are difficult to see in this choice, as discussed below). In this
construction, each additional flavor corresponds to an additional
semi-infinite D5-brane. As discussed in the previous section, each
such additional brane adds one real parameter to the theory, which it
is natural to interpret as the (real) mass of the corresponding quark
hypermultiplet (which is in the $\bf N_c$ representation).

\fig{$SU(4)$ gauge theory with two (massive) flavors in five dimensions.}
{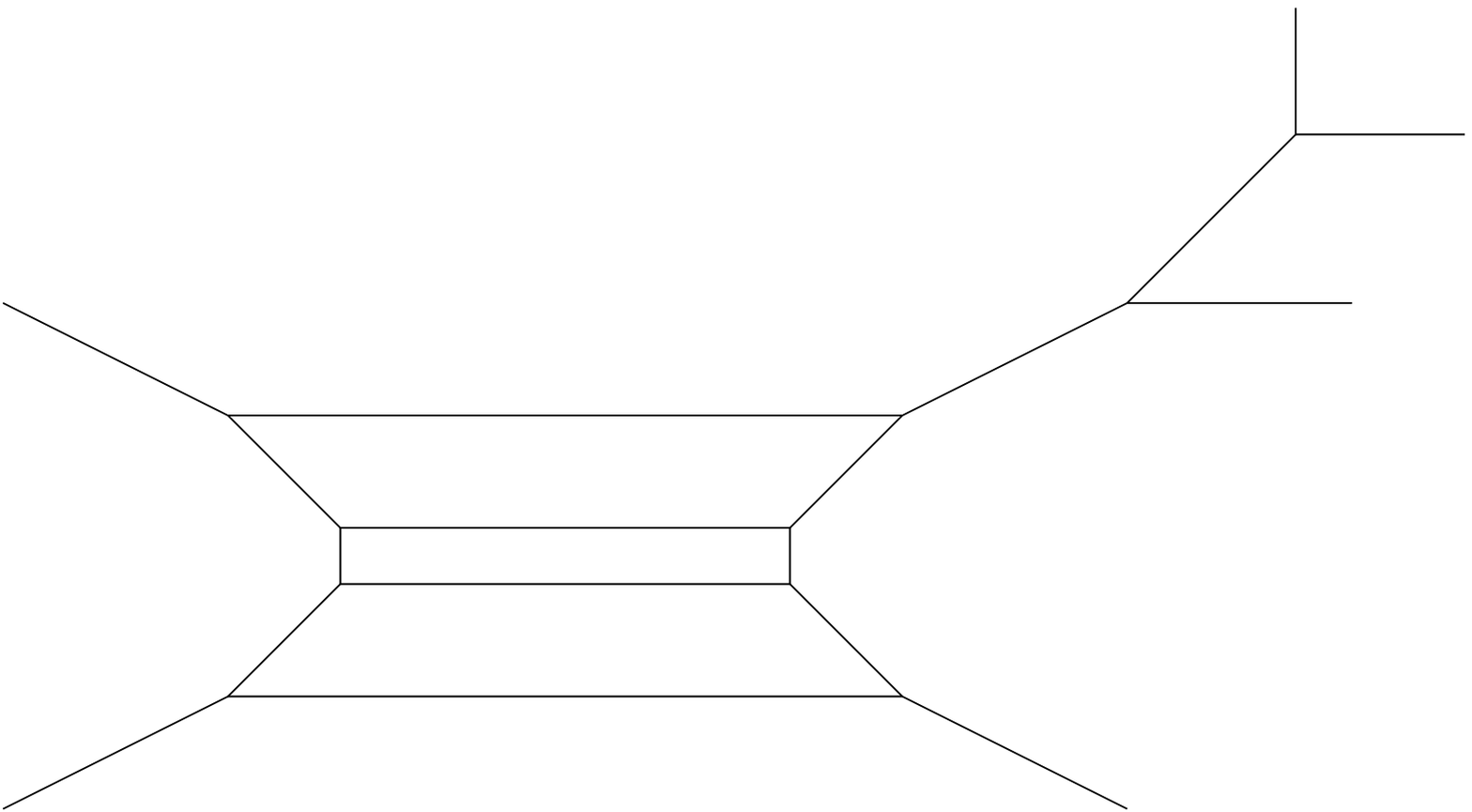}{10 truecm}
\figlabel\SUfour

An example of this type of constructions appears in figure \SUfour.
Note that for $N_c > 2$ and odd values of $N_f$, these theories have a
$\IZ_2$ anomaly \fived\ and cannot be properly defined as 5D gauge
theories. The brane constructions of these theories, however, seem to
be consistent, so the anomaly is probably cancelled by higher
dimensional effects (as in \refs{\ghm,\bdl}), but we have not checked
this carefully.

As we did for the pure gauge theories, we can compute the monopole
tensions by looking at areas in the brane construction, and we
find agreement with the known field theory results. 

\fig{$SU(2)$ gauge theory with one massless quark at infinite
coupling (and finite $\phi$), and with four massless quarks at finite
coupling (and finite $\phi$).}{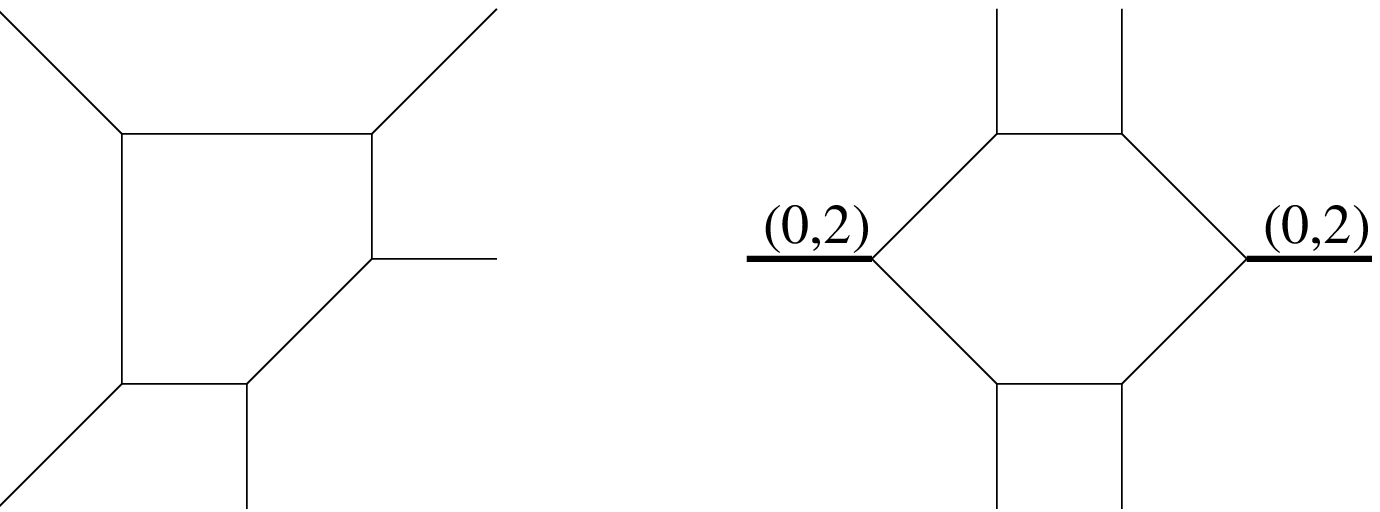}{12 truecm}
\figlabel\SUtwonew

The simplest examples are $SU(2)$ gauge theories, two of which (with
massless quarks) are drawn in figure \SUtwonew. In the first example,
corresponding to $N_f=1$ at the strong coupling fixed point, we find a
tension of ${7\over 8} \phi^2$ (with the same normalization we used
for the pure gauge theory), as expected since the coefficient of the
$\phi^2$ term should be proportional to $8-N_f$. In the second
example, which is $N_f=4$ at finite coupling, we find a tension of
$1/g_0^2 \phi + {1\over 2} \phi^2$, again as expected. Note that in
this case only
an $SU(2)\times SU(2)$ non-Abelian factor of the global symmetry is
visible in the brane construction, even though we expect to have
an $SO(8)$ global flavor symmetry at low energies.

As we did for the pure gauge theories, we can take the gauge coupling to
infinity and find strong coupling fixed points at the origin of moduli
space. With $N_f$ flavors, these fixed points correspond to an
intersection of $4+N_f$ half 5-branes. As before, it is not clear if
fixed points having the same Coulomb branch but different asymptotic
5-branes are the same or not. In a similar way, we can also construct
$SU(N_c)$ theories by using $N_c$ parallel $(p,q)$ 5-branes, and again
it is not clear if the fixed points related to these are the same as
the previous ones or not. Brane configurations corresponding to
$SO$ and $USp$ gauge theories (analyzed in \fived) may be constructed
in a similar way by adding orientifolds (as in \refs{\oogvaf,\ejs})
but will not be discussed here.

\fig{An attempt to construct the $SU(2)$ theory with $N_f=5$.}
{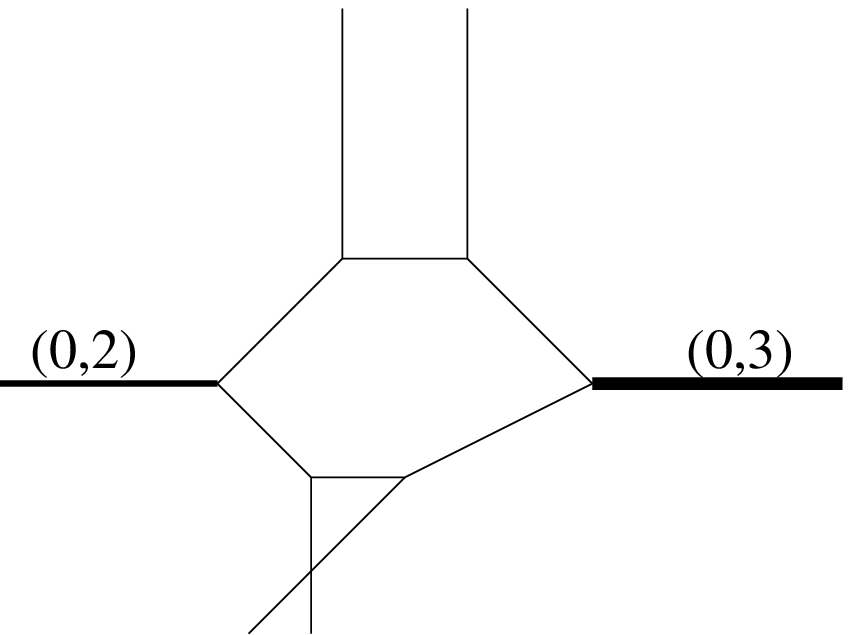}{8 truecm}
\figlabel\SUtwofive

Brane constructions of the type discussed above exist if (and only if)
$N_f \leq 2N_c$. If we try to add more flavors, we find that we must
have more intersections of 5-branes than the ones we had before along
the Coulomb branch, as in figure \SUtwofive. It is then natural to
assume that the field theory is no longer just the $SU(N_c)$ theory
with $N_f$ flavors, which does not have a positive definite metric for
$N_f > 2N_c$ and $N_c \geq 3$ \fived, but that there are additional
degrees of freedom corresponding to the additional finite branes in
the construction. For $N_c=2$ consistent field theories are argued
to exist also for $N_f=5,6,7,8$ \seibergfive, but we do not know how to see
this in the brane construction (in which $N_c=2$ does not appear to be
special). This may be related to the exceptional global symmetries of
the strong coupling fixed points of these theories, though the global
symmetries are not always visible in the brane construction. In any
case, the strong coupling fixed points exist in the brane construction
(as defined above)
for any values of $N_f$ and $N_c$, though for $N_f > 2N_c$ they can no
longer be deformed into $SU(N_c)$ gauge
theories. We conjecture that these fixed points correspond to new
superconformal field theories.

So far we have not discussed the Higgs branches of the gauge theories
we were considering, which should exist for $N_f \geq 2$. These
branches are, in fact, difficult to see when the flavors come from
semi-infinite 5-branes (and also in the analogous situation of
\newwitten). Starting from the standard construction of quark flavors
in \hw, going to the constructions we describe involves moving branes
away to infinity in the $x_7$ direction ($x_6$ in the notation of
\hw), and the Higgs branches apparently go off to infinity as
well. Note that we expect Higgs branches in our construction to be
related to moving branes in the $x_{4,5,6}$ directions, since the
$SU(2)_R$ symmetry should be broken along them. Since none of the
finite 5-brane segments of our construction can move in these
directions, it seems that the Higgs branch is, in fact, related to
the semi-infinite 5-branes, which have low-energy 6 dimensional field
theories. The relation between the five dimensional low-energy physics
and the six dimensional
physics is, therefore, more complicated here (and in \newwitten) than
in \hw, and we will not discuss it further here.

\subsec{Reproducing the $SU(2)$ flows}
\subseclab{\flows}

In this section we present an application of the brane constructions
to an analysis of the $SU(2)$ gauge theories with $N_f \leq 1$. The
strong coupling fixed point of the $N_f=1$ theory (called the $E_2$
theory) has two
deformations, which may be interpreted as the quark mass and the gauge
coupling. The analysis of these deformations was performed in
\refs{\morsei,\dkv}, where it was found that one can flow to
two different fixed points of the $N_f=0$ theory, which were denoted
$E_1$ and $\tilde{E_1}$, and that one can flow from the latter to an
$E_0$ fixed point (which has a one dimensional Coulomb branch but no
gauge theory interpretation).

\fig{The $E_2$ fixed point and some flows in parameter space away 
from it.}
{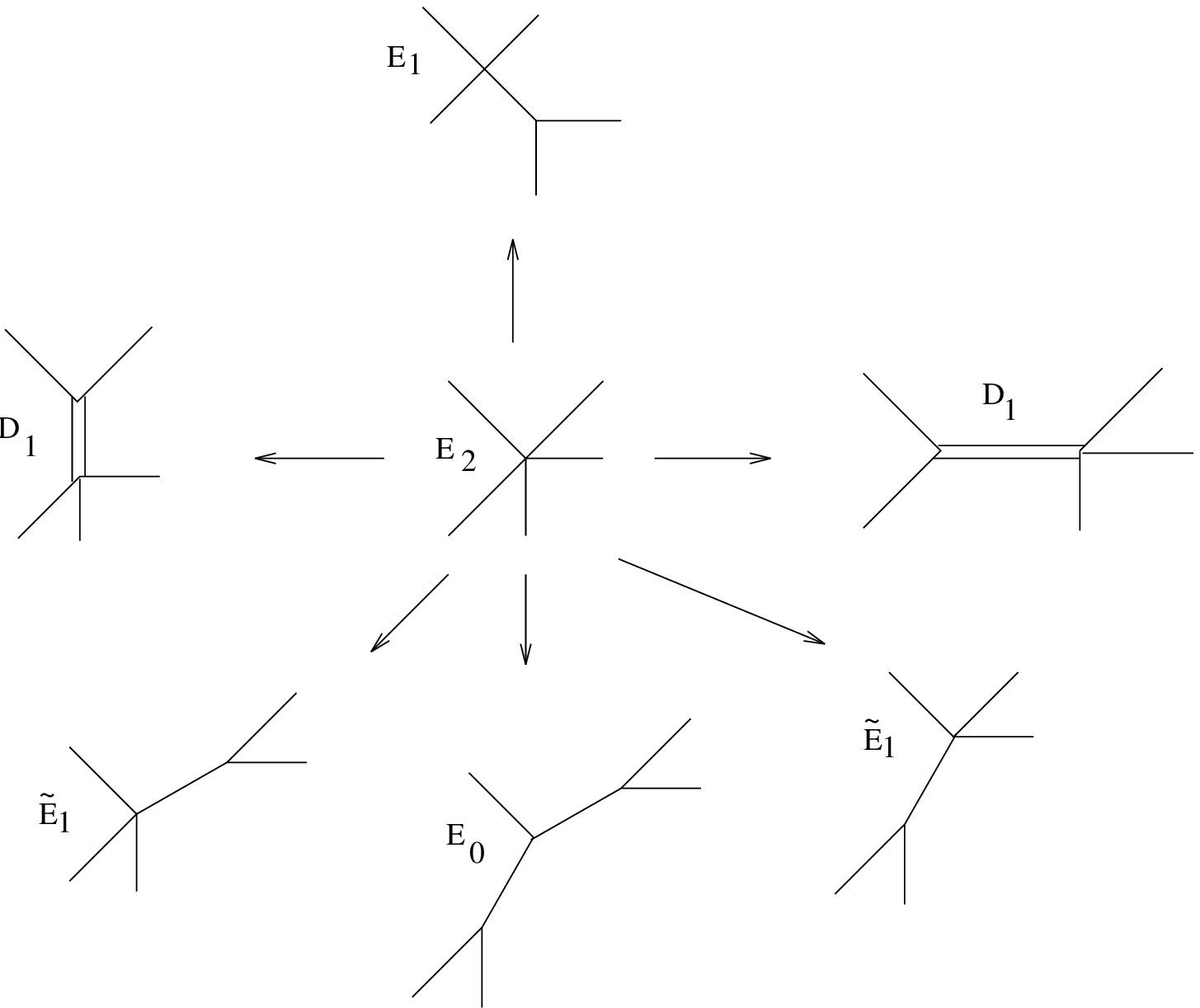}{15 truecm}
\figlabel\Etwoflows

All of these results may be rederived in the brane construction,
as described in figure \Etwoflows, which is the brane realization of
figure 1 in \morsei. 
The $E_1$ and $\tilde{E_1}$ theories are both special cases of the
strong coupling fixed points of the pure $SU(2)$ gauge theory
described in \S3.1, with different asymptotic branes. The $E_0$ theory
is realized as an intersection of three half 5-branes, so it has
(using the results of
\S3.2) no parameters, as
expected \morsei. The three 5-branes may be chosen
to have $(p_i,q_i) = \{(1,1),(-2,1),(1,-2)\}$.

\fig{The Coulomb branch of the $E_0$ theory.}
{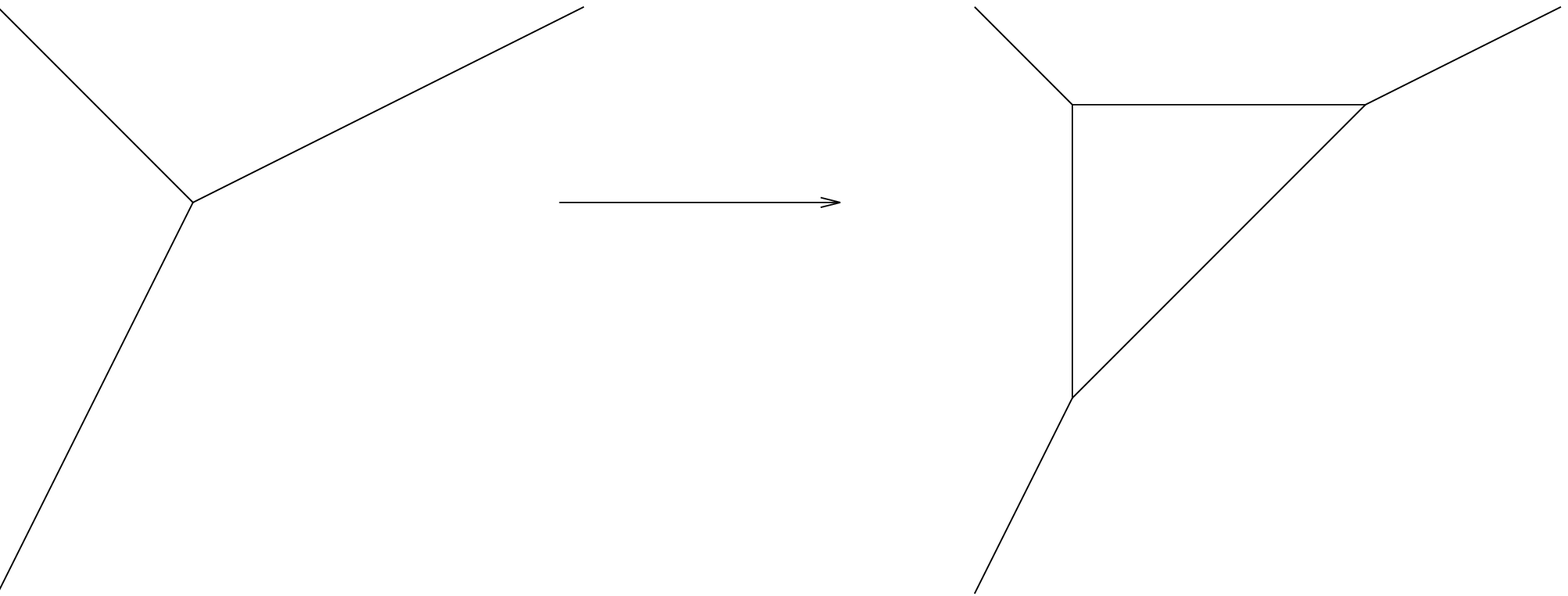}{10 truecm}
\figlabel\EzeroCoul

For most intersections of three half 5-branes there is no Coulomb
branch, but in this case there is a Coulomb branch emanating from the
$E_0$ point, as shown in figure \EzeroCoul. The other intersection
points in figure \Etwoflows\ all have no Coulomb branches emanating
from them, and appear to correspond to trivial theories.

Generic fixed points, as described in \S3.2, will probably not
correspond in any obvious way to gauge theories, like the $E_0$
theory. It is not clear if in general they can also be reached by
flows from the $SU(N)$ fixed points. In any case, we are able to
analyze their deformations and the flows between them, as described in
the previous sections.

\subsec{$SU(n)\times SU(m)$ gauge groups}

Other field theory results we can easily reconstruct in the brane
picture pertain to products of gauge groups. It was argued in \fived\
that products of simple gauge groups (with matter charged under both
groups) cannot have strong coupling fixed points, since the effective
coupling for one factor of the gauge group would become negative along
the Coulomb branch corresponding to the other group. 

\fig{The $SU(2)\times SU(2)$ theory. Figure (a) shows a point on the
Coulomb branch, while figure (b) shows what happens when we go along
the Coulomb branch of one of the groups and not the other.}
{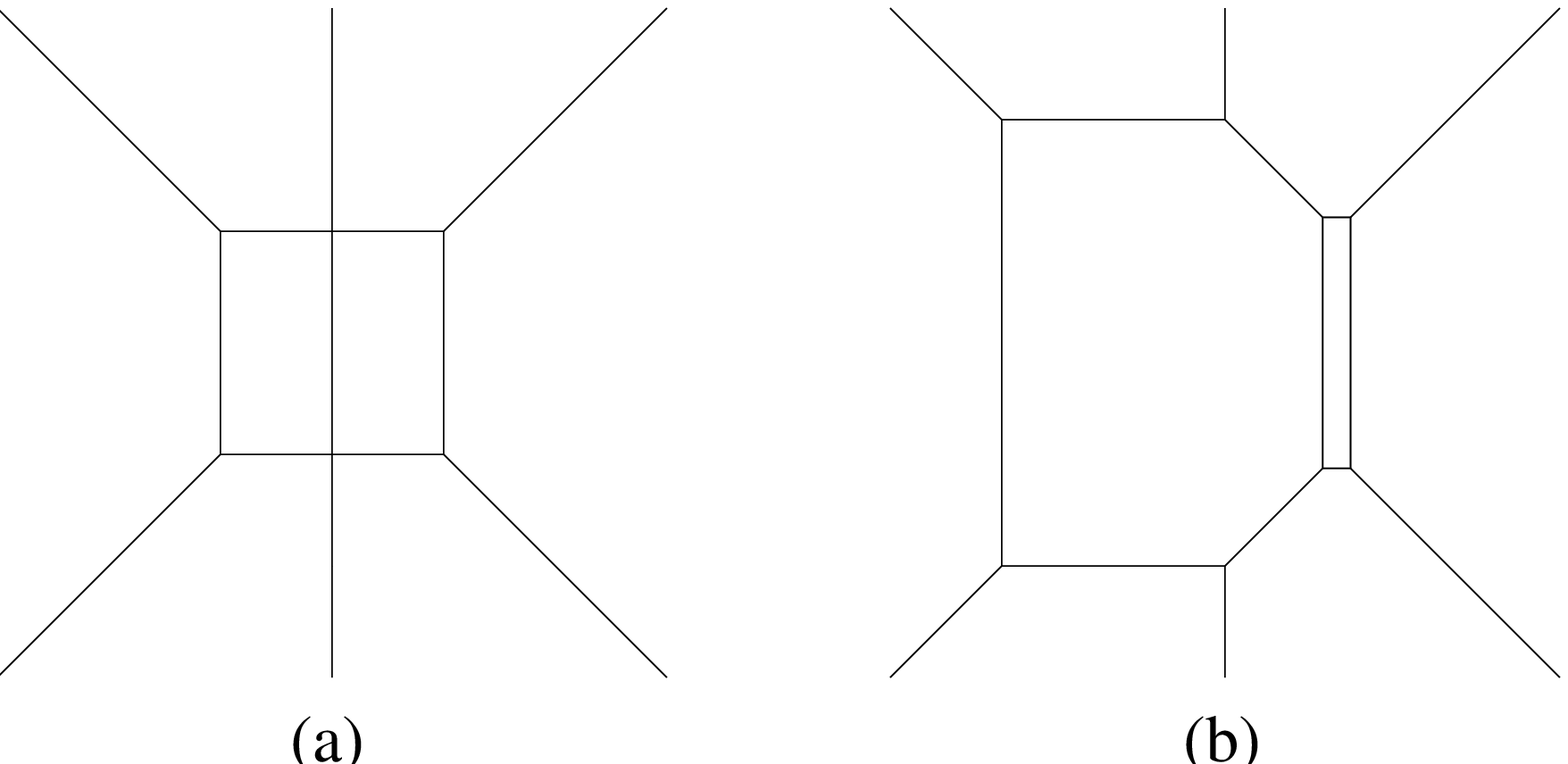}{10 truecm}
\figlabel\Sutwosutwo

How do we see this in the brane constructions ? These are a
generalization of the constructions analyzed in \newwitten, and an
example for $SU(2)\times SU(2)$ is described in figure
\Sutwosutwo. Naively, the theory in figure \Sutwosutwo(a) is an
$SU(2)\times SU(2)$ gauge theory with a hypermultiplet in the $(2,2)$
representation, at a point on the Coulomb branch emanating from this
theory's infinite coupling fixed point. However, we see in figure
\Sutwosutwo(b) that if we try to go far along the Coulomb branch of
one of the $SU(2)$ groups and not of the other, the description of the
other $SU(2)$ group breaks down. In fact, we seem to get its
continuation beyond infinite coupling, as we discussed in
\S3.1. The brane configuration continues to describe some low-energy
field theory, but we can no longer describe it simply as a
$SU(2)\times SU(2)$ field theory.

In fact, the $SU(2)\times SU(2)$ theory described here appears very
similar to the $SU(3)$ theory with $N_f=2$ massless quarks, which is
of the type described in \S3.3. This is what we get if we turn the
figure by 90 degrees, and replace NS branes with D branes\foot{Note
that the configurations we discuss in this section are not invariant
under general rotations in the $x_3-x_7$ plane, since our choice of
the unbroken supersymmetry is not consistent with such rotations, but
rotations by 90 degrees are equivalent to the S generator of
$SL(2,\IZ)$ as far as the unbroken supersymmetry is concerned, so they
are expected not to change the strong coupling fixed points.}. Thus,
the strong coupling fixed point that we naively attributed to the
$SU(2)\times SU(2)$ theory is really the strong coupling fixed point
of the $SU(3)$ theory with $N_f=2$.  The $SU(3)$ description makes
sense on the whole Coulomb branch, while the $SU(2)\times SU(2)$
description does not. It should be interesting to understand the exact
relationship between these two theories, and the generalization of
this statement to arbitrary gauge groups.

\newsec{From 5D Fixed Points to 3D Fixed Points}

In this section we combine the methods of the previous two sections,
and relate the 5D $N=1$ superconformal theories constructed in \S3 to
3D $N=2$ superconformal theories. As in \S2, this is done by
stretching finite 3-branes in the 0126 directions between 5-branes,
but now instead of using just NS 5-branes to bound the 3-branes, we
will use ``polymeric'' 5-branes of the type considered in the previous
section. Since the branes we use are a subset of the ones described in
\S2, this configuration preserves $1/8$ of the supersymmetry,
corresponding to $N=2$ in three dimensions.

Consider any configuration of 5-branes like those described in the
previous section, filling the 01289 directions and corresponding to a
``polymer'' in the $x_3-x_7$ plane. A 3-brane in the 0126 directions
can end on any $(p,q)$ 5-brane, so, in particular, it can end at any
point on the ``polymer'' and have flat directions corresponding to
moving along the ``polymer''. In order to have a 3D theory we want the
3-brane to be finite in the $x_6$ direction, and if we want to
preserve the flat directions corresponding to moving along the
``polymer'' we need to have two copies of the same ``polymer'' at two
different values of $x_6$. Now, if we stretch a 3-brane (or several
3-branes) between these ``polymers'' we get a 3D $N=2$ theory, with
flat directions corresponding to moving along the ``polymer''. 

When the 3-brane is at a generic point on the ``polymer'' it ends (on
both sides) on some $(p,q)$ 5-brane, with $p$ and $q$ relatively
prime\foot{A 3-brane ending on a 5-brane with $p$ and $q$ not
relatively prime was conjectured in \egk\ to be described by a
$\tr(X^k)$ superpotential for an adjoint field $X$, but we will not
discuss this here.}. This configuration is
$SL(2,\IZ)$-dual to a 3-brane ending on both sides on a NS 5-brane,
which is described (at low energies) by the $U(1)$ $N=4$ gauge theory,
as discussed in
\hw. The massless fields are thus in a $(p,q)$-vector multiplet, which
is related by electric-magnetic duality in the D3-brane to the
standard vector multiplet (for a D3-brane between two D5-branes, this
multiplet may be described as a hypermultiplet, as in \hw, but its
description in terms of a vector multiplet is more natural here).
The motion along the $(p,q)$ 5-brane (along the ``polymer'' and in the
$x_8$ and $x_9$ directions) corresponds to the Coulomb branch
of this theory. From the $N=2$ point of view there is one $(p,q)$
vector multiplet and one chiral multiplet, describing the $x_{89}$
position of the 3-brane, which will exist (and be decoupled) in all
configurations of this type. For $N_c$ D3-branes together, all these
fields are enhanced to adjoints of $U(N_c)$.

Things get more interesting when the 3-brane approaches some
intersection of 5-branes. Along each branch separately, there is a
massless $(p,q)$ vector 
multiplet. The vector multiplets corresponding to
5-branes with different values of $(p,q)$ are not mutually local (they
are
related by $SL(2,\IZ)$ transformations in the field theory of the
3-brane). At the intersection point we thus expect to get some
non-trivial superconformal field theory, at least when we have more
than one D3-brane so that there are charged particles (W bosons) under
all the mutually non-local vector multiplets (as in similar four
dimensional cases with $N=2$ \refs{\argdoug,\apsw} and $N=1$ \aks\
supersymmetry). Each non-trivial 5D SCFT described in the previous
section thus gives rise to a 3D $N=2$ SCFT as well. In fact, we get an
infinite series of such SCFTs, one for every value of $N_c$. 

Even trivial 5D constructions of the type described in \S3 give rise
to interesting 3D $N=2$ SCFTs. For instance, the field theory
corresponding to a D3-brane moving on the ``polymer'' described in
figure \NSD\ has three Coulomb-like branches, which are all non-local
with respect to each other. Along each Coulomb branch the low-energy
theory is an $N=4$ theory, but with different $SU(2)_R\times SU(2)_L$
global symmetries (one of the $SU(2)$ factors corresponds to rotations
in $x_8,x_9$ and the direction in the $x_3-x_7$ plane along the
``polymer,'' while the other corresponds to rotations in $x_4,x_5$ and
the direction in the $x_3-x_7$ plane orthogonal to the
``polymer''). The breaking of $N=4$ to $N=2$ is felt at low energies
only at the intersection point of the Coulomb branches.

A possible generalization of this construction still utilizes the same
``polymers'' on both sides, but constructed on one side from branes
spanning the 01245 direction, and on the other side from branes
spanning the 01289 direction (as well as the directions corresponding
to the ``polymer''). The supersymmetry in the 3D theory is then still
$N=2$, and the only difference from the previous case is that the
scalar corresponding to the $x_{89}$ position of the D3-brane is now
massive, so along the Coulomb branches we have the pure $N=2$ SYM
theory.

Another possible generalization is to theories with two different
``polymers'' on both sides of the D3-branes, but with some of their
``line segments'' still overlapping (in the $x_3-x_7$ plane). For
instance, we could have the configuration of figure \NSD\ on one side,
and the same configuration shifted up in $x_3$ on the other side. In
this case, the Coulomb branch corresponding to the D3-brane moving
along the NS 5-brane still remains, but it ends when the NS 5-brane
ends on one side, and there seem to be no other branches emanating
from this end point. Presumably, such end points also correspond to
interesting 3D $N=2$ superconformal field theories.

Finally, we would like to discuss the implications of our discussion
here for the conjecture of \bh\ regarding chiral symmetry in the brane
construction. For the 3D $N=2$ theories, this conjecture is that when
$N_f$ D5-branes overlap (in $4+1$ dimensions) with a NS 5-brane, there
is actually an $SU(N_f)\times SU(N_f)$ gauge symmetry on the 5-branes,
which is a chiral symmetry for the D3-branes ending on these
D5-branes. Following our discussion of section 3, the intersection of
the $N_f$ D5-branes with the NS 5-brane would seem to be some
superconformal theory (though the 5D part of this theory may be
trivial, since there is no Coulomb branch coming out of the
intersection point), and it is not obvious if it can really be
described in terms of an $SU(N_f)\times SU(N_f)$ gauge symmetry.

\fig{The 5-brane configuration conjectured to correspond to chiral
symmetry, and a deformation of this configuration which splits the
D5-branes.}
{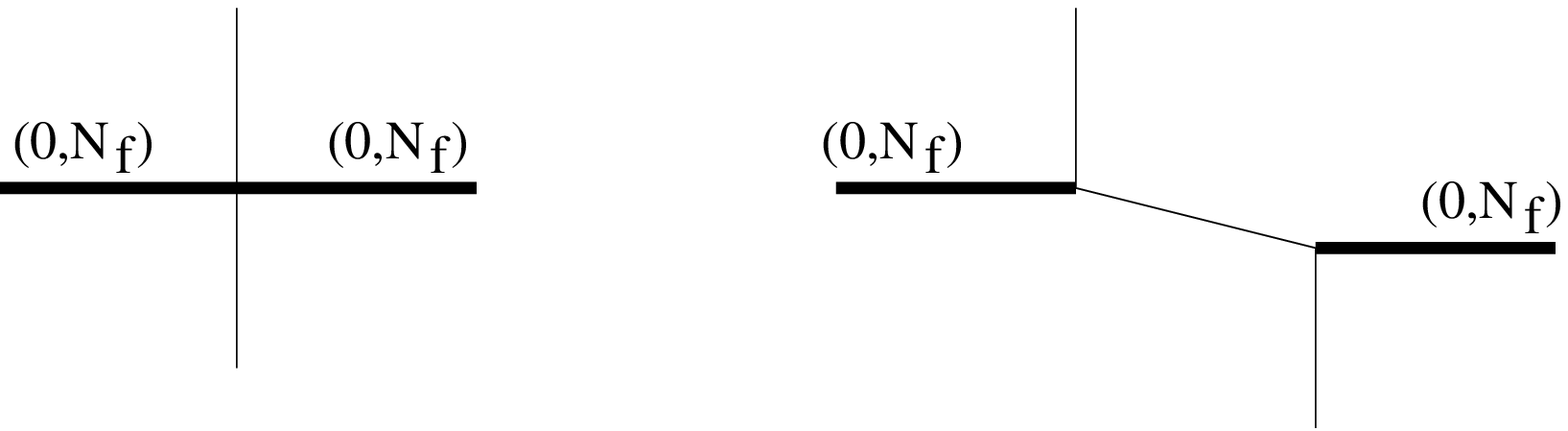}{13 truecm}
\figlabel\Split

However, unlike the 4D case discussed in \bh, here we can split the
$N_f$ D5-branes along the NS 5-brane, as shown in figure \Split. Then,
it seems clear that there is indeed an $SU(N_f)\times SU(N_f)$ gauge
symmetry in the $5+1$ dimensional gauge theory of the
D5-branes. However, it is no longer clear what field theory
corresponds to a D3-brane ending on this configuration. As discussed
above, the theory of a D3-brane ending at the intersection point
between the NS 5-brane and the $N_f$ D5-branes corresponds to some
superconformal fixed point, which does not seem to correspond to a
gauge theory with $N_f$ massless quarks (and no anti-quarks), as
suggested by \bh\ (though it might be some deformation of this
theory). The issue of chiral symmetry in the brane configurations
deserves further investigation.

\vskip 0.5in

\centerline{\bf Acknowledgments}

We would like to thank John Brodie, Kenneth Intriligator, David
Morrison, Hiroshi Ooguri, Yaron Oz, Nathan Seiberg, and especially
Edward Witten for useful discussions. The work of O.A. was
supported in part by DOE grant DE-FG02-96ER40559.  The research of
A.H. is supported in part by NSF grant PHY-9513835.

\vskip 0.5in

\centerline {\bf Note Added}

As this paper was being completed, we received \egkrs, which has some
overlap with our discussion in section 2.

\listrefs

\end